\documentclass[10pt]{iopart}
\usepackage{iopams}  
\usepackage{xcolor}
\usepackage{graphicx}
\usepackage{stackrel}
\usepackage{longtable}
\usepackage{url}

\usepackage{amsopn}
\usepackage[numbers]{natbib}
\DeclareMathOperator*{\argmin}{arg\,min}

\newcommand{\order}{\ensuremath{\mathcal{O}}}
\usepackage{booktabs}

\begin{document}

\topical[Bridging observation, theory and numerical simulation of the ocean using ML]{Bridging observation, theory and numerical simulation of the ocean using Machine Learning}

\author{Maike Sonnewald$^{1,2,3}$\footnote{Present address:
Princeton University, Program in Atmospheric and Oceanic Sciences, 300 Forrestal Rd., Princeton, NJ 08540}, Redouane Lguensat$^{4,5}$, Daniel C. Jones$^6$, Peter D. Dueben$^7$, Julien Brajard$^{5,8}$, V. Balaji$^{1,2,4}$
}
 \ead{maikes@princeton.edu}

\address{$^1$Princeton University, Program in Atmospheric and Oceanic Sciences, Princeton, NJ 08540, USA}
\address{$^2$NOAA/OAR Geophysical Fluid Dynamics Laboratory, Ocean and Cryosphere Division, Princeton, NJ 08540, USA}
\address{$^3$University of Washington, School of Oceanography, Seattle, WA, USA}
\address{$^4$ Laboratoire des Sciences du Climat et de l'Environnement (LSCE-IPSL), CEA Saclay, Gif Sur Yvette, France}
\address{$^5$ LOCEAN-IPSL, Sorbonne Universit\'e, Paris, France}
\address{$^6$British Antarctic Survey, NERC, UKRI, Cambridge, UK}
\address{$^7$European Centre for Medium Range Weather Forecasts, Reading, UK}
\address{$^8$ Nansen Center (NERSC),  Bergen, Norway}

\vspace{10pt}
\begin{indented}
\item[]June 2021
\end{indented}
\begin{abstract}

Progress within physical oceanography has been concurrent with the increasing sophistication of tools available for its study. The incorporation of machine learning (ML) techniques offers exciting possibilities for advancing the capacity and speed of established methods and for making substantial and serendipitous discoveries. Beyond vast amounts of complex data ubiquitous in many modern scientific fields, the study of the ocean poses a combination of unique challenges that ML can help address. The observational data available is largely spatially sparse, limited to the surface, and with few time series spanning more than a handful of decades. Important timescales span seconds to millennia, with strong scale interactions and numerical modeling efforts complicated by details such as coastlines. This review covers the current scientific insight offered by applying ML and points to where there is imminent potential. We cover the main three branches of the field: observations, theory, and numerical modeling. Highlighting both challenges and opportunities, we discuss both the historical context and salient ML tools. We focus on the use of ML in situ sampling and satellite observations, and the extent to which ML applications can advance theoretical oceanographic exploration, as well as aid numerical simulations. Applications that are also covered include model error and bias correction and current and potential use within data assimilation. While not without risk, there is great interest in the potential benefits of oceanographic ML applications; this review caters to this interest within the research community.

\end{abstract}

%
\vspace{2pc}
\submitto{\ERL}
%
%
\ioptwocol

\section{Introduction}
\label{sec:intro}

\subsection{Oceanography: observations, theory, and numerical simulation}
\label{sec:oceans}


The physics of the oceans have been of crucial importance, curiosity and interest since prehistoric times, and today remain an essential element in our understanding of weather and climate, and a key driver of biogeochemistry and overall marine resources. The eras of progress within oceanography have gone hand in hand with the tools available for its study. Here, the current progress and potential future role of machine learning (ML) techniques is reviewed and briefly put into historical context. ML adoption is not without risk, but is here put forward as having the potential to accelerate scientific insight, performing tasks better and faster, along with allowing avenues of serendipitous discovery. This review focuses on physical oceanography, but concepts discussed are applicable across oceanography and beyond.

Perhaps the principal interest in oceanography was originally that of navigation, for exploration, commercial and military purposes. Knowledge of the ocean as a dynamical entity with predictable features-- the regularity of its currents and tides -- must have been known for millennia. Knowledge of oceanography likely helped the successful colonization of Oceania \cite{Montenegro201612426}, and similarly Viking and Inuit navigation \cite{Haine2008}, the oldest known dock was constructed in Lothal with knowledge of the tides dating back to 2500--1500 BCE\cite{cartwright2007}, and Abu Ma'shar of Baghdad in the 8th century CE correctly attributed the existence of tides to the Moon's pull. 

The ocean measurement era, determining temperature and salinity at depth from ships, starts in the late 18th century CE. While the tools for a theory of the ocean circulations started to become available in the early 19th century CE with the Navier-Stokes equation, observations remained at the core of oceanographic discovery. The first modern oceanographic textbook was published in 1855 by M. Mauri, whose work in oceanography and politics served the slave trade across the Atlantic, around the same time CO$_2$'s role in climate was recognized \cite{Foote1856, Tyndall1859}. The first major global observational synthesis of the ocean can be traced to the Challenger expeditions of 1873-75 CE \cite{ref:deacon2018}, where observational data from various areas was brought together to gain insight into the global ocean. The observational synthesis from the Challenger expeditions gave a first look at the global distribution of temperature and salinity including at depth, revealing the 3-dimensional structure of the ocean. 

Quantifying the time mean ocean circulation remains challenging, as ocean circulation features strong local and instantaneous fluctuations. Improvements in measurement techniques allowed the Swedish oceanographer Ekman to elucidate the nature of the wind-driven boundary layer \cite{Ekman1905}. Ekman used observations taken on an expedition led by the Norwegian oceanographer and explorer Nansen, where the Fram was intentionally frozen into the Arctic ice. The ``dynamic method'' was introduced by Swedish oceanographer Sandström and the Norwegian oceanographer Helland-Hansen \cite{SHH1903}, allowing the indirect computation of ocean currents from density estimates under the assumption of a largely laminar flow. This theory was developed further by Norwegian meteorologist Bjerknes into the concept of \textit{geostrophy}, from the Greek geo for earth and strophe for turning. This theory was put to the test in the extensive Meteor expedition in the Atlantic from 1925-27 CE; they uncovered a view of the horizontal and vertical ocean structure and circulation that is strikingly similar to our present view of the Atlantic meridional overturning circulation \cite{MW1923, Richardson2008}. 

While the origins of Geophysical Fluid Dynamics (GFD) can be traced back to Laplace or Archimedes, the era of modern GFD can be seen to stem from linearizing the Navier-Stokes equations, which enabled progress in understanding meteorology and atmospheric circulation. For the ocean, pioneering dynamicists include Sverdrup, Stommel, and Munk, whose theoretical work still has relevance today \cite{Stommel1948, Munk1950}. As compared to the atmosphere, the ocean circulation exhibits variability over a much larger range of timescales, as noted by \cite{ref:munk1950}, likely spanning thousands of years rather than the few decades of detailed ocean observations available at the time. 
Yet, there are phenomena at intermediate timescales (that is, months to years) which seemed to involve both atmosphere and ocean, e.g \cite{ref:namias1959}, and indeed Sverdrup suggests the importance of the coupled atmosphere-ocean system in \cite{ref:sverdrup1942}. In the 1940s much progress within GFD was also driven by the second world war (WWII). The introduction of accurate navigation through radar introduced with WWII worked a revolution for observational oceanography together with bathythermographs intensively used for submarine detection. Beyond \textit{in situ} observations, the launch of Sputnik, the first artificial satellite, in 1957 heralded the era of ocean observations from satellites. Seasat, launched on the 27th of June 1978, was the first satellite dedicated to ocean observation. 


Oceanography remains a subject that must be understood with an appreciation of available tools, both observational and theoretical, but also numerical. While numerical GFD can be traced back to the early 1900s \cite{abbe1901, bjerknes1904, richardson2007}, it became practical with the advent of numerical computing in the late 1940s, complementing that of the elegant deduction and more heuristic methods that one could call ``pattern recognition'' that had prevailed before \cite{ref:balaji2021}. The first ocean general circulation model with specified global geometry were developed by Bryan and Cox \cite{bryan1968nonlinear,bryan1997numerical} using finite-difference methods. This work paved the way for what now is a major component of contemporary oceanography. The first coupled ocean-atmosphere model of \cite{ref:manabebryan1969} eventually led to their use for studies of the coupled Earth system, including its changing climate. The low-power integrated circuit that gave rise to computers in the 1970s also revolutionized observational oceanography, enabling instruments to reliably record autonomously. This has enabled instruments such as moored current meters and profilers, drifters, and floats through to hydrographic and velocity profiling devices that gave rise to microstructure measurements. Of note is the fleet of free-drifting Argo floats, beginning in 2002, which give an extraordinary global dataset of profiles \cite{ref:roemmichetal2009}. Data assimilation (DA) is the important branch of modern oceanography combining what is often sparse observational data with either numerical or statistical ocean models to produce observationally-constrained estimates with no gaps. Such estimates are referred to as an 'ocean state', which is especially important for understanding locations and times with no available observations.

Together the innovations within observations, theory, and numerical models have produced distinctly different pictures of the ocean as a dynamical system, revealing it as an intrinsically turbulent and topographically influenced circulation \cite{Wunsch2002, Garabato2012}. Key large scale features of the circulation depend on very small scale phenomena, which for a typical model resolution remain parameterized rather than explicitly calculated. For instance, fully accounting for the subtropical wind-driven gyre circulation and associated western boundary currents relies on an understanding of the vertical transport of vorticity input by the wind and output at the sea floor, which is intimately linked to mesoscale (ca. 100km) flow interactions with topography \cite{hughes2001western,eden2010western}. It has become apparent that localized small-scale turbulence (0-100km) can also impact the larger-scale, time-mean overturning and lateral circulation by affecting how the upper ocean interacts with the atmosphere \cite{thomas2008ocean,ferrari2008parameterization,hazeleger2000eddy}. The prominent role of the small scales on the large scale circulation has important implications for understanding the ocean in a climate context, and its representation still hinges on the further development of our fundamental understanding, observational capacity, and advances in numerical approaches.

\begin{figure*}
  \includegraphics[width=\textwidth]{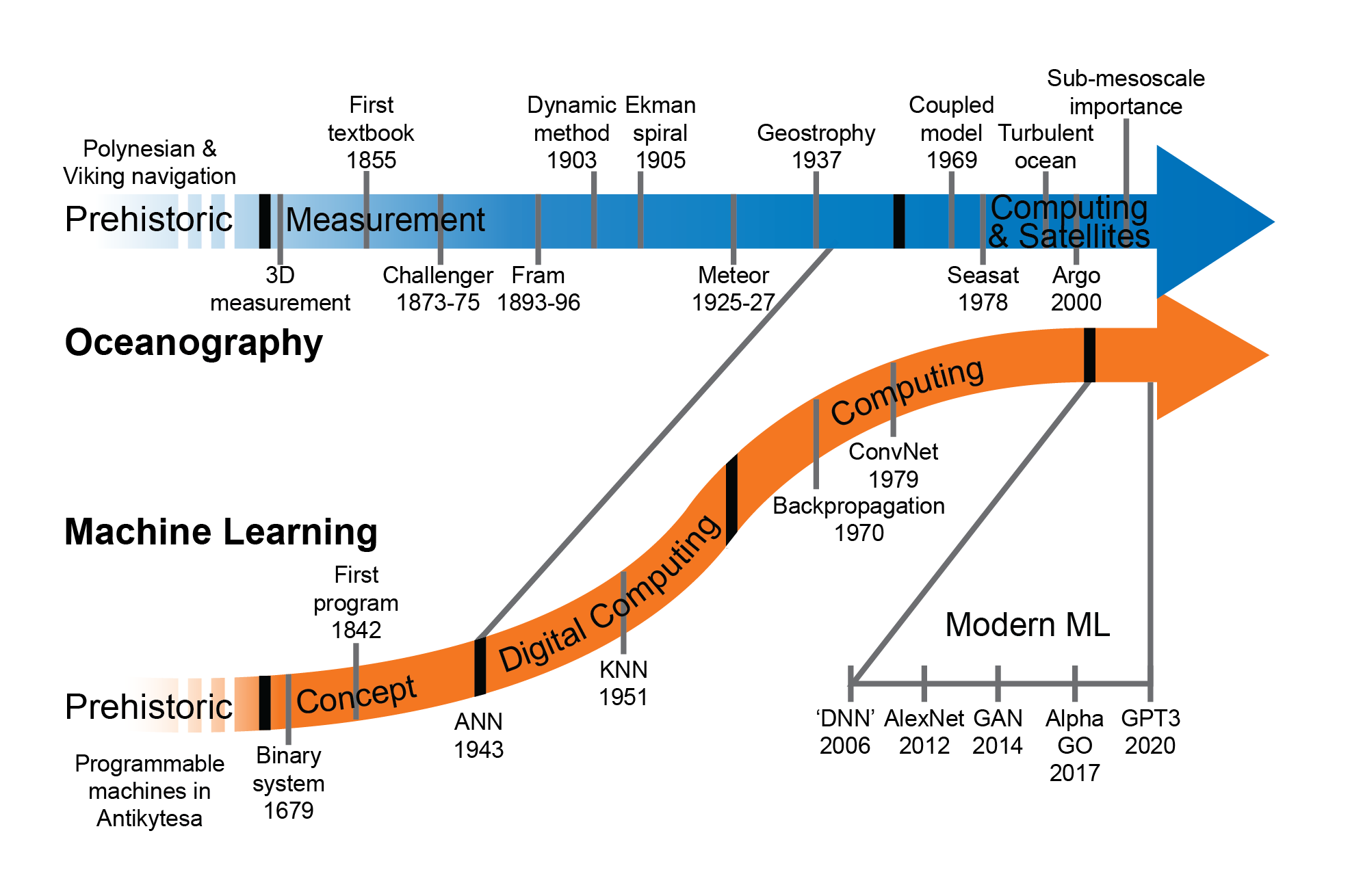}
  \caption{\textbf{Timeline sketch of oceanography (blue) and ML (orange)}. The timelines of oceanography and ML are moving towards each other, and interactions between the fields where ML tool as are incorporated into oceanography has the potential to accelerate discovery in the future. Distinct `events' marked in grey. Each field has gone through stages (black), with progress that can be attributed to the available tools. With the advent of computing, the fields were moving closer together in the sense that ML methods generally are more directly applicable. Modern ML is seeing an very fast increase in innovation, with much potential for adoption by oceanographers. See table \ref{tabl3} for acronyms.}
  \label{fig:arrows}
\end{figure*}

The development of both modern oceanography and ML techniques have happened concurrently, as illustrated in Fig. \ref{fig:arrows}. This review summarizes the current state of the art in ML applications for physical oceanography and points towards exciting future avenues. We wish to highlight certain areas where the emerging techniques emanating from the domain of ML demonstrate potential to be transformative. ML methods are also being used in closely-related fields such as atmospheric science. However, within oceanography one is faced with a unique set of challenges rooted in the lack of long-term and spatially dense data coverage. While in recent years the surface of the ocean is becoming well observed, there is still a considerable problem due to sparse data, particularly in the deep ocean. Temporally, the ocean operates on timescales from seconds to millennia, and very few long term time series exist. There is also considerable scale-interaction, which also necessitates more comprehensive observations.

There remains a healthy skepticism towards some ML applications, and calls for ``trustworthy'' ML are also coming forth from both the European Union and the United States government (Assessment List for Trustworthy Artificial Intelligence [ALTAI], and mandate E.O. 13960 of Dec 3, 2020). Within the physical sciences and beyond, trust can be fostered through transparency. For ML, this means moving beyond the ``black box'' approach for certain applications. Moving away from this black box approach and adopting a more transparent approach involves gaining insight into the learned mechanisms that gave rise to ML predictive skill. This is facilitated by either building a priori \textit{interpretable} ML applications or by retrospectively \textit{explaining} the source of predictive skill, coined interpretable and explainable artificial intelligence (IAI and XAI, respectively \cite{Rudin2019,irrgang2021,Beucler2021,Sonnewald2021}). An example of interpretability could be looking for coherent structures (or ``clusters'') within a closed budget where all terms are accounted for. Explainability comes from, for example, tracing the weights within a Neural Network (NN) to determine what input features gave rise to its prediction. With such insights from transparent ML, a synthesis between theoretical and observational branches of oceanography could be possible. Traditionally, theoretical models tend towards oversimplification, while data can be overwhelmingly complicated. For advancement in the fundamental understanding of ocean physics, ML is ideally placed to identify salient features in the data that are comprehensible to the human brain. With this approach, ML could significantly facilitate a generalization beyond the limits of data, letting data reveal possible structural errors in theory. With such insight, a hierarchy of conceptual models of ocean structure and circulation could be developed, signifying an important advance in our understanding of the ocean.

\begin{figure*}
  \includegraphics[width=\textwidth]{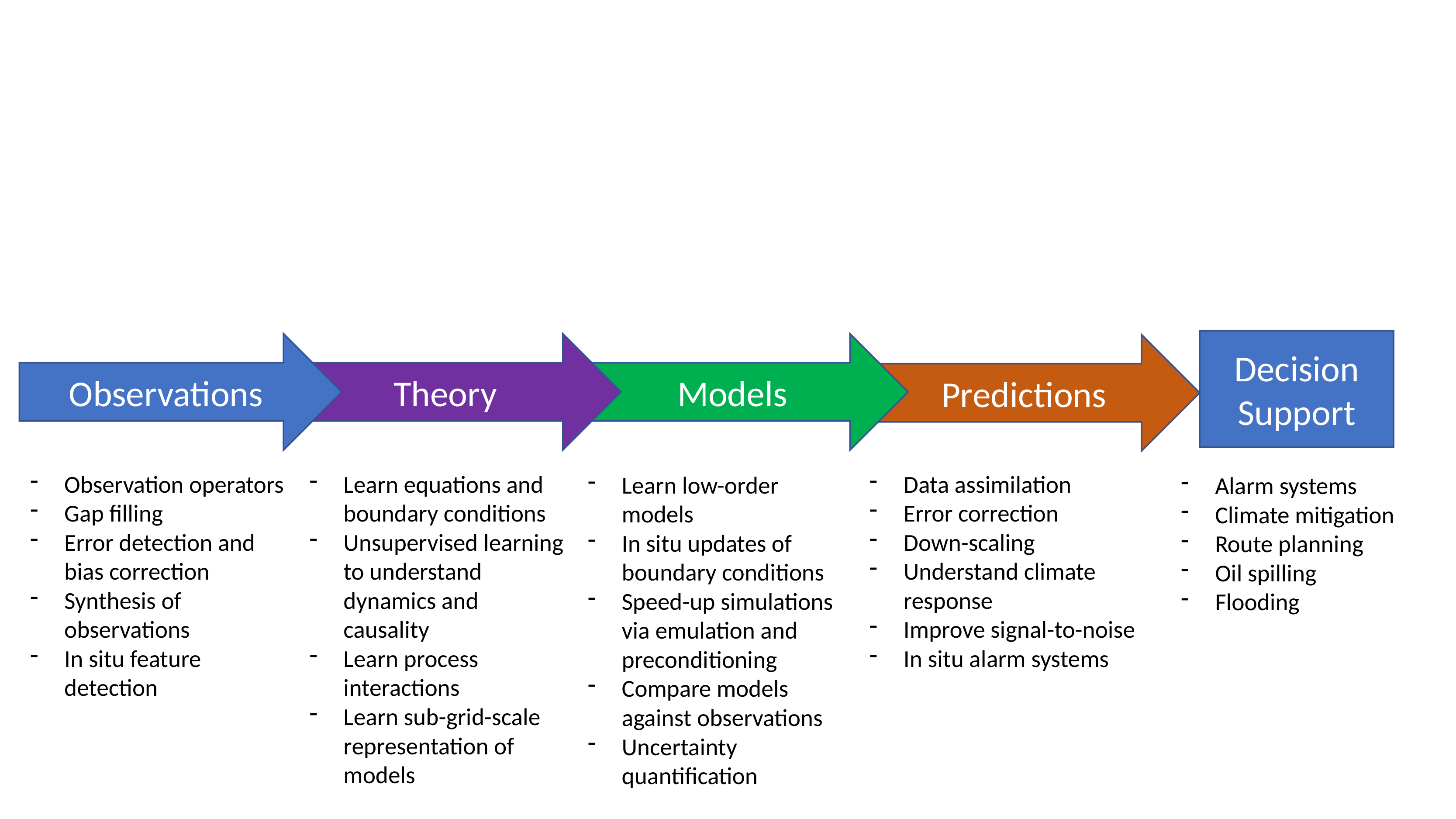}
  \caption{\textbf{Machine learning within the components of oceanography.} A diagram capturing the general flow of knowledge, highlighting the components covered in this review. Separating the categories (arrows) is artificial, with ubiquitous feed-backs between most components, but serves as an illustration.}
  \label{fig:ml4ocean}
\end{figure*}

In this review, we introduce ML concepts (Section~\ref{sec:introML}), and some of its current roles in the atmospheric and Earth System Sciences (Section~\ref{sec:mlESS}), highlighting particular areas of note for ocean applications. The review follows the structure outline illustrated in Fig. \ref{fig:ml4ocean}, with the ample overlap noted through cross referencing the text. We review ocean observations (Section~\ref{sec:obs}), sparsely observed for much history, but now yielding increasingly clear insight into the ocean and its 3D structure. In Section~\ref{sec:theory} we examine a potential synergy between ML and theory, with the intent to distill expressions of theoretical understanding by dataset analysis from both numerical and observational efforts. We then progress from theory to models, and the encoding of theory and observations in numerical models (Section~\ref{sec:models}). We highlight some issues involved with ML-based prediction efforts (Section~\ref{sec:predict}), and end with a discussion of challenges and opportunities for ML in the ocean sciences (Section~\ref{sec:synthesis}). These challenges and opportunities include the need for transparent ML, ways to support decision makers and a general outlook. Appendix \ref{tabl3} has a list of acronyms.

\subsection{Concepts in ML}
\label{sec:introML}
Throughout this article, we will mention some concepts from the ML literature. We find it then natural to start this paper with a brief introduction to some of the main ideas that shaped the field of ML.

ML, a sub-domain of Artificial Intelligence (AI), is the science of providing mathematical algorithms and computational tools to machines, allowing them to perform selected tasks by ``learning" from data. This field has undergone a series of impressive breakthroughs over the last years thanks to the increasing availability of data and the recent developments in computational and data storage capabilities. Several classes of algorithms are associated with the different applications of ML. They can be categorized into three main classes: supervised learning, unsupervised learning, and reinforcement learning (RL). In this review, we focus on the first two classes which are the most commonly used to date in the ocean sciences.

\subsubsection{Supervised learning}

Supervised learning refers to the task of inferring a relationship between a set of inputs and their corresponding outputs. In order to establish this relationship, a ``labeled" dataset is used to constrain the learning process and assess the performance of the ML algorithm. Given a dataset of $N$ pairs of input-output training examples $\{(x^{(i)},y^{(i)})\}_{i \in 1..N}$ and a loss function $\mathcal{L}$ that represents the discrepancy between the ML model prediction and the actual outputs, the parameters $\boldsymbol{\theta}$ of the ML model $f$ are found by solving the following optimization problem:
\begin{equation}
    \boldsymbol{\theta}^{*}=\stackrel[\boldsymbol{\theta}]{}{\argmin} \frac{1}{N} \sum_{i=1}^{N} \mathcal{L}\left(f\left(x^{(i)} ; \boldsymbol{\theta}\right), y^{(i)}\right).
    \label{eq:ml-minimize}
\end{equation}

If the loss function is differentiable, then gradient descent based algorithms can be used to solve equation~\ref{eq:ml-minimize}. These methods rely on an iterative tuning of the models' parameters in the direction of the negative gradient of the loss function. At each iteration $k$, the parameters are updated as follows:
\begin{equation}
\mathbf{\boldsymbol{\theta}}_{k+1}=\mathbf{\boldsymbol{\theta}}_{k}-\mu \nabla \mathcal{L}\left(\mathbf{\boldsymbol{\theta}}_{k}\right),
\end{equation}
where $\mu$ is the rate associated with the descent and is called the learning rate and $\nabla$ the gradient operator.

Two important applications of supervised learning are regression and classification. Popular statistical techniques such as Least Squares or Ridge Regression, which have been around for a long time, are special cases of a popular supervised learning technique called Linear Regression (in a sense, we may consider a large number of oceanographers to be early ML practitioners.) For regression problems, we aim to infer continuous outputs and usually use the mean squared error (MSE) or the mean absolute error (MAE) to assess the performance of the regression. In contrast, for supervised classification problems we sort the inputs to a number of classes or categories that have been pre-defined. In practice, we often transform the categories into probability values of belonging to some class and use distribution-based distances such as the cross-entropy to evaluate the performance of the classification algorithm.

Numerous types of supervised ML algorithms have been used in the context of ocean research, as detailed in the following sections. Notable methods include: 
\begin{itemize}
    \item \textit{Linear univariate (or multivariate) regression (LR)}, where the output is a linear combination of some explanatory input variables. LR is one of the first ML algorithms to be studied extensively and used for its ease of optimization and its simple statistical properties \cite{montgomery2021introduction}. 
    \item \textit{k-Nearest Neighbors (KNN)}, where we consider an input vector, find its $k$ closest points with regard to a specified metric, then classify it by a plurality vote of these $k$ points. For regression, we usually take the average of the values of the $k$ neighbors. KNN is also known as ``analog methods" in the numerical weather prediction community \cite{lorenz1969atmospheric}.
    \item \textit{Support Vector Machines (SVM)} \cite{cortes1995support}, where the classification is done by finding a linear separating hyperplane with the maximal margin between two classes (the term ``margin" here denotes the space between the hyperplane and the nearest points in either class.) In case of data which cannot be separated linearly, the use of the \textit{kernel trick} projects the data into a higher dimension where the linear separation can be done. \textit{Support Vector Regression (SVR)} are an adaption of SVMs for regression problems.
    \item \textit{Random Forests (RF)} that are a composition of a multitude of Decision Trees (DT). DTs are constructed as a tree-like composition of simple decision rules \cite{biau2016random}.
    \item \textit{Gaussian Process Regression (GPR)} \cite{williams1996gaussian}, also called kriging, is a general form of the optimal interpolation algorithm, which has been used in the oceanographic community for a number of years 
    \item \textit{Neural Networks (NN)}, a powerful class of universal approximators that are based on compositions of interconnected nodes applying geometric transformations (called affine transformations) to inputs and a nonlinearity function called an ``activation function" \cite{cybenko1989approximation}
\end{itemize}

The recent ML revolution, i.e. the so-called Deep Learning (DL) era that began in the early 2010s, sparked off thanks to the scientific and engineering breakthroughs in training neural networks (NN), combined with the proliferation of data sources and the increasing computational power and storage capacities. The simplest example of this advancement is the efficient use of the algorithm of backpropagation (known in the geocience community as the adjoint method) combined with stochastic gradient descent for the training of multi-layer NNs, i.e. NNs with multiple layers, where each layer takes the result of the previous layer as an input, applies the mathematical transformations and then yields an input for the next layer \cite{goodfellow2016deep}. DL research is a field receiving intense focus and fast progress through its use both commercially and scientifically, resulting in new types of "architectures" of NNs, each adapted to particular classes of data (text, images, time series, etc.) \cite{schmidhuber2015deep,lecun2015deep}. We briefly introduce the most popular architectures used in deep learning research and highlight some applications:
\begin{itemize}
    \item Multilayer Perceptrons (MLP): when used without qualification, this term refers to fully connected feed forward multilayered neural networks. They are composed of an input layer that takes the input data, multiple hidden layers that convey the information in a "feed forward" way (i.e. from input to output with no exchange backwards), and finally an output layer that yields the predictions. Any neuron in a MLP is connected to all the neurons in the previous and to those of next layer, thus the use of the term "fully connected". MLPs are mostly used for tabular data.
    \item Convolutional Neural Networks (ConvNet): contrarily to MLPs, ConvNets are designed to take into account the local structure of particular type of data such as text in 1D, images in 2D, volumetric images in 3D, and also hyperspectral data such as that used in remote sensing. Inspired by the animal visual cortex, neurons in ConvNets are not fully connected, instead they receive information from a subarea spanned by the previous layer called the "receptive field". In general, a ConvNet is a feed forward architecture composed of a series of convolutional layers and pooling layers and might also be combined with MLPs. A convolution is the application of a filter to an input that results in an activation. One convolutional layer consist of a group of "filters" that perform mathematical discrete convolution operations, the result of these convolutions are called "feature maps". The filters along with biases are the parameters of the ConvNet that are learned through backpropagation and stochastic gradient descent. Pooling layers serve to reduce the resolution of feature maps which lead to compressing the information and speeding up the training of the ConvNet, they also help the ConvNet become invariant to small shift in input images \cite{lecun2015deep}. ConvNets benefited much from the advancements in GPU computing and showed great success in the computer vision community. 
    \item Recurrent Neural Networks (RNN): with an aim to model sequential data such as temporal signals or text data, RNNs were developed with a hidden state that stores information about the history of the sequences presented to its inputs. While theoretically attractive, RNNs were practically found to be hard to train due to the exploding/vanishing gradient problems, i.e. backpropagated gradients tend to either increase too much or shrink too much at each time step\cite{hochreiter1991}. Long Short Term Memory (LSTM) architecture provided a solution to this problem through the use of special hidden units \cite{schmidhuber2015deep}. LSTMs are to date the most popular RNN architectures and are used in several applications such as translation, text generation, time series forecasting, etc. Note that a variant for spatiotemporal data was developed to integrate the use of convolutional layers, this is called ConvLSTM \cite{shi2015convolutional}.
\end{itemize}

\subsubsection{Unsupervised learning}
Unsupervised learning is another major class of ML. In these applications, the datasets are typically unlabelled. The goal is then to discover patterns in the data that can be used to solve particular problems. One way to say this is that unsupervised classification algorithms identify sub-populations in data distributions, allowing users to identify structures and potential relationships among a set of inputs (which are sometimes called ``features" in ML language). Unsupervised learning is somewhat closer to what humans expect from an intelligent algorithm, as it aims to identify latent \textit{representations} in the structure of the data while filtering out unstructured noise. 
At the NeurIPS 2016 conference, Yann LeCun, a DL pioneer researcher, highlighted the importance of unsupervised learning using his \textit{cake analogy}: "If machine learning is a cake, then unsupervised learning is the actual cake, supervised learning is the icing, and RL is the cherry on the top."

Unsupervised learning is achieving considerable success in both clustering and dimensionality reduction applications. Some of the unsupervised techniques that are mentioned throughout this review are:

\begin{itemize}
    \item \textit{k-means}, a popular and simple space-partitioning clustering algorithm that finds classes in a dataset by minimizing within-cluster variances \cite{steinhaus1956division}. Gaussian Mixture Models (GMMs) can be seen as a generalization of the k-means algorithm that assumes the data can be represented by a mixture (i.e. linear combination) of a number of multi-dimensional Gaussian distributions \cite{mclachlan1988mixture}. 
    \item Kohonen maps [also called \textit{Self Organizing Maps} (SOM)] is a NN based clustering algorithm that leverages topology of the data; nearby locations in a learned map are placed in the same class \cite{kohonen1982self}. K-means can be seen as a special case of SOM with no information about the neighborhood of clusters. 
    \item \textit{t-SNE} and \textit{UMAP} are two other clustering algorithms which are often used for not only finding clusters but also because of their data visualization properties which enables a two or three dimensional graphical rendition of the data \cite{van2008visualizing, mcinnes2018umap}. These methods are useful for representing the structure of a high-dimensional dataset in a small number of dimensions that can be plotted. For the projection, they use a measure of the ``distance'' or ``metric'' between points, which is a sub-field of mathematics where methods are increasingly implemented for t-SNE or UMAP. 
    \item \textit{Principal Component Analysis} (PCA) \cite{obukhov1947statistically}, the simplest and most popular dimensionality reduction algorithm. Another term for PCA is Empirical Orthogonal Function analysis (EOF), which has been used by physical oceanographers for many years, also called Proper Orthogonal Decomposition (POD) in computational fluids literature.
    \item \textit{Autoencoders} (AE) are NN-based dimensionality reduction algorithms, consisting of a bottleneck-like architecture that learns to reconstruct the input by minimzing the error between the output and the input (i.e. ideally the data given as input and output of the autoencoder should be interchangeable). A central layer with a lower dimension than the original inputs' dimension is called a ``code" and represents a compressed representation of the input \cite{kramer1991nonlinear}. 
    \item \textit{Generative modeling}: a powerful paradigm that learns the latent features and distributions of a dataset and then proceeds to generate new samples that are plausible enough to belong to the initial dataset. Variational Auto-encoders (VAEs) and Generative Adversarial Networks (GANS) are two popular techniques of generative modeling that benefited much from the DL revolution \cite{kingma2013auto,goodfellow2014generative}.
\end{itemize}

Between supervised and unsupervised learning lies \textit{semi-supervised learning}. It is a special case where one has access to both labeled and unlabeled data. A classical example is when labeling is expensive, leading to a small percentage of labeled data and a high percentage of unlabeled data.

Reinforcement learning is the third paradigm of ML; it is based on the idea of creating algorithms where an \textit{agent} explores an \textit{environment} with the aim of reaching some goal. The agent learns through a trial and error mechanism, where it performs an \textit{action} and receives a response (reward or punishment), the agent learns by maximizing the expected sum of rewards \cite{szepesvari2010algorithms}. The DL revolution did also affect this field and led to the creation of a new field called deep reinforcement learning (Deep RL) \cite{sutton2018reinforcement}. A popular example of Deep RL that got huge media attention is the algorithm AlphaGo developed by DeepMind which beat human champions in the game of Go \cite{silver2017mastering}.

The importance of understanding why an ML method arrived at a result is not confined to oceanographic applications. Unsupervised ML lends itself more readily to being interpreted (IAI), but for example for methods building on DL or NN in general, a growing family of methods collectively referred to as Additive Feature Attribution (AFA) is becoming popular, largely applied for XAI. AFA methods aim to explain predictive skill retrospectively. These methods include connection weight approaches, Local Interpretable Model-agnostic Explanations (LIME), Shapley Additive Explanation (SHAP) and Layer-wise Relevance Propagation (LRP) \cite{Olden2004, Lapuschkin2015, lime, shap, toms_physically_2020, Beucler2021, Sonnewald2021, Montavon2018}. Non-AFA methods rooted in `saliency' mapping also exist \cite{mcgovern_making_2019}.

The goal of this review paper is not to delve into the definitions of ML techniques but only to briefly introduce them to the reader and recommend references for further investigation. The textbook by Christopher Bishop \cite{bishop2006pattern} covers essentials of the fields of pattern recognition and ML. William Hsieh's book \cite{hsieh2009machine} is probably one of earliest attempts at writing a comprehensive review of ML methods targeted at earth scientists. Another notable review of statistical methods for physical oceanography is the paper by Wikle et al. \cite{wikle2013modern}. We also refer the interested reader to the book of Goodfellow et al. \cite{goodfellow2016deep} to learn more about the theoretical foundations of DL and some of its applications in science and engineering.
 

\subsection{ML in atmospheric and the wider Earth system sciences}
\label{sec:mlESS}


Precursors to modern ML methods, such as regression and principal component analysis, have of course been used in many fields of Earth system science for decades. The use of PCA, for example, was popularized in meteorology in \cite{lorenz1956empirical}, as a method of dimensionality reduction of large geospatial datasets, where Lorenz also speculates here on the possibility of purely statistical methods of long-term weather prediction based on a representation of data using PCA. Methods for discovering correlations and links, including possible causal links, between dataset features using formal methods have seen much use in Earth system science. e.g \cite{ref:barnettpreisendorfer1987}. For example, Walker \cite{ref:walker1928} was tasked with discovering the cause for the interannual fluctuation of the Indian monsoon, whose failure meant widespread drought in India, and in colonial times also famine \cite{ref:davis2001}. To find possible correlations, Walker put to work an army of Indian clerks to carry out a vast computation by hand across all available data. This led to the discovery of the Southern Oscillation, the seesaw in the West-East temperature gradient in the Pacific, which we know now by its modern name, El Ni{\~n}o Southern Oscillation (ENSO). Beyond observed correlations, theories of ENSO and its emergence from coupled atmosphere-ocean dynamics appeared decades later \cite{ref:zebiakcane1987}. Walker speaks of statistical methods of discovering ``weather connections in distant parts of the earth'', or teleconnections. The ENSO-monsoon teleconnection remains a key element in diagnosis and prediction of the Indian monsoon \cite{ref:swapnaetal2014}, \cite{ref:swapnaetal2018}. These and other data-driven methods of the pre-ML era are surveyed in \cite{ref:brethertonetal1992}. ML-based predictive methods targeted at ENSO are also being established \cite{ham2019deep}. Here, the learning is not directly from observations but from models and reanalysis data, and outperform some dynamical models in forecasting ENSO.

There is an interplay between data-driven methods and physics-driven methods that both strive to create insight into many complex systems, where the ocean and the wider Earth system science are examples. As an example of physics-driven methods \cite{ref:balaji2021}, Bjerknes and other pioneers discussed in Section~\ref{sec:oceans} formulated accurate theories of the general circulation that were put into practice for forecasting with the advent of digital computing. Advances in numerical methods led to the first practical physics-based atmospheric forecast \cite{phillips1956general}. Until that time, forecasting often used data-driven methods ``that were neither algorithmic nor based on the laws of physics'' \cite{ref:nebeker1995}. ML offers avenues to a synthesis of data-driven and physics-driven methods. In recent years, as outlined below in Section~\ref{sec:compute}, new processors and architectures within computing have allowed much progress within forecasting and numerical modeling overall. ML methods are poised to allow Earth system science modellers to increase the efficient use of modern hardware even further. It should be noted however that ``classical'' methods of forecasting such as analogues also have become more computationally feasible, and demonstrate equivalent skill, e.g \cite{ref:dingetal2019}. The search for analogues has become more computationally tractable as well, although there may also be limits here \cite{ref:dool1994}.


Advances in numerical modeling brought in additional understanding of elements in Earth system science which are difficult to derive, or represent from first principles. Examples include cloud microphysics or interactions with the land surface and biosphere. For capturing cloud processes within models, the actual processes governing clouds take place at scales too fine to model and will remain out of reach of computing for the foreseeable future \cite{ref:schneideretal2017}. A practical solution to this is finding a representation of the aggregate behavior of clouds at the resolution of a model grid cell. This has proved quite difficult and progress over many decades has been halting \cite{ref:bonyetal2013}. 
The use of ML in deriving representations of clouds is now an entire field of its own. Early results include the results of \cite{ref:gentineetal2018}, using NNs to emulate a ``super-parameterized'' model. In the super-parameterized model, there is a clear (albeit artificial) separation of scales between the ``cloud scale'' and the large scale flow. When this scale separation assumption is relaxed, some of the stability problems associated with ML re-emerge \cite{brenowitz2018prognostic}. There is also a fundamental issue of whether learned relationships respect basic physical constraints, such as conservation laws \cite{ref:lingetal2016}. Recent advances (\cite{ref:yuvaletal2021}, \cite{ref:beucleretal2021}) focus on formulating the problem in a basis where invariances are automatically maintained. But this still remains a challenge in cases where the physics is not fully understood.

There are at least two major efforts for the systematic use of ML methods to constrain the cloud model representations in GCMs. First, the calibrate-emulate-sample (CES \cite{cleary2021,ref:dunbaretal2020}) approach uses a more conventional model for a broad calibration of parameters also referred to as ``tuning''\cite{ref:hourdinetal2017}. This is followed by an emulator, that calibrates further and quantifies uncertainties. The emulator is an ML-based model that reproduces most of the variability of the reference model, but at a lower computational cost. The low computational cost enables the emulator to be used to produce a large ensemble of simulations, that would have been too computationally expensive to produce using the model that the emulator is based on. It is important to retain the uncertainty quantification aspect (represented by the emulated ensemble) in the ML context, as it is likely that the data in a chaotic system only imperfectly constrain the loss function. Second, emulators can be used to eliminate implausible parameters from a calibration process, demonstrated by the HighTune project \cite{ref:couvreuxetal2020, ref:hourdinetal2020}. This process can also identify ``structural error'', indicating that the model formulation itself is incorrect, when no parameter choices can yield a plausible solution. Model errors are discussed in Section~\ref{sec:bias}. In an ocean context, the methods discussed here can be a challenge due to the necessary forwards model component. Note also, that ML algorithms such as GPR are ubiquitous in emulating problems thanks to their built-in uncertainty quantification. GPR methods are also popular because their application involves a low number of training samples, and function as inexpensive substitutes for a forward model. 

Model resolution that is inadequate for many practical purposes has led to the development of data-driven methods of ``downscaling''. For example climate change adaptation decision-making at the local level based on climate simulations too coarse to feature enough detail. Most often, a coarse-resolution model output is mapped onto a high-resolution reference truth, for example given by observations \cite{Vandal2018,Adewoyin2021}. Empirical-statistical downscaling (ESD, \cite{ref:benestadetal2008}) is an example of such methods. While ESD emphasized the downscaling aspect, all of these downscaling methods include a substantial element of bias correction. This is highlighted in the name of some of the popular methods such as Bias Correction and Spatial Downscaling \cite{ref:woodetal2004} and Bias Corrected Constructed Analogue \cite{ref:maureretal2010}. These are trend‐preserving statistical downscaling algorithms, that combine bias correction with the analogue method of Lorenz (1969)\cite{ref:lorenz1969}. ML methods are rapidly coming to dominate the field as discussed in Section~\ref{sec:bias}, with examples ranging from precipitation (e.g \cite{ref:vandaletal2019}), surface winds and solar outputs \cite{Stengel2020}, as well as to unresolved river transport \cite{ref:ghoshmujumdar2008}. Downscaling methods continue to make the assumption that transfer functions learned from present-day climate continue to hold in the future. This stationarity assumption is a potential weakness of data-driven methods (\cite{o2018using, ref:dixonetal2016}), that requires a synthesis of data-driven and physics-based methods as well.

\section{Ocean observations}
\label{sec:obs}
Observations continue to be key to oceanographic progress, with ML increasingly being recognised as a tool that can enable and enhance what can be learned from observational data, performing conventional tasks better/faster, as well as bring together different forms of observations, facilitating comparison with model results. ML offers many exciting opportunities for use with observations, some of which are covered in this section and in section \ref{sec:predict} as supporting predictions and decision support.

The onset of the satellite observation era brought with it the availability of a large volume of effectively global data, challenging the research community to use and analyze this unprecedented data stream. Applications of ML intended to develop more accurate satellite-driven products go back to the 90's~\cite{thiria1993neural}. These early developments were driven by the data availability, distributed in normative format by the space agencies, and also by the fact that models describing the data were either empirical (e.g. marine biogeochemistry~\cite{schiller1999neural}) or too computationally costly and complex (e.g. radiative transfer~\cite{key1998tools}). More recently, ML algorithms have been used to fuse several satellite products~\cite{guimbard2012smos} and also satellite and in-situ data~\cite{mustapha2014automatic,chapman_reconstruction_2017, martinez2020reconstructing, kavanaugh_seascapes_2016, denvil2019lsce}. For the processing of satellite data, ML has proven to be a valuable tool for extracting geophysical information from remotely sensed data (e.g. \cite{duncan_distinctiveness_2019, Castellani2006}), whereas a risk of using only conventional tools is to exploit only a more limited subset of the mass of data available. These applications are based mostly on instantaneous or very short-term relationships and do not address the problem of how these products can be used to improve our ability to understand and forecast the oceanic system. Further use for current reconstruction using ML \cite{Manucharyan2021}, heat fluxes \cite{George2019}, the 3-dimensional circulation\cite{Sonnewald2021}, and ocean heat content\cite{Irrgang2019} are also being explored.

There is also an increasingly rich body of literature mining ocean in-situ observations. These leverage a range of data, including Argo data, to study a range of ocean phenomena. Examples include assessing North Atlantic mixed layers \cite{Maze2017}, describing spatial variability in the Southern Ocean \cite{Jones2019}, detecting El Niño events \cite{Houghton2020}, assessing how North Atlantic circulation shifts impacting heat content \cite{Desbruyeres2021}, and finding mixing hot spots \cite{Rosso2020}. ML has also been successfully applied to ocean biogeochemistry. While not covered in  detail here, examples include mapping oxygen \cite{giglio2018} and CO$_2$ fluxes \cite{watson2020,Landschutzer2014,Bushinsky2019}.

Modern in-situ classification efforts are often property-driven, carrying on long traditions within physical oceanography. For example, characteristic groups or ``clusters'' of salinity, temperature, density or potential vorticity have typically been used to delineate important water masses and to assess their spatial extent, movement, and mixing \cite{HellandHansen1916, HanawaTalley2001}. However, conventional identification/classification techniques assume that these properties stay fixed over time. The techniques largely do not take interannual and longer timescale variability into account. The prescribed ranges used to define water masses are often somewhat ad-hoc and specific (e.g. mode waters are often tied to very restrictive density ranges) and do not generalize well between basins or across longer timescales \cite{Aoki2020}. Although conventional identification/classification techniques will continue to be useful well into the future, unsupervised ML offers a robust, alternative approach for objectively identifying structures in oceanographic observations \cite{Jones2019, Rosso2020,Pauthenet2019, Boehme2021}.   

To analyze data, dimensionality and noise reduction methods have a long history within oceanography. PCA is one such method, which has had a profound influence on oceanography since Lorenz first introduced it to the geosciences in 1956 \cite{lorenz1956empirical}. Despite the method's shortcomings related to strong statistical assumptions and misleading applications, it remains a popular approach \cite{Monahan2009}. PCA can be seen as a super sparse rendition of k-means clustering \cite{DingHe2004} with the assumption of an underlying normal distribution in its commonly used form. Overall, different forms of ML can offer excellent advantages over more commonly used techniques. For example, many clustering algorithms can be used to reduce dimensionality according to how many significant clusters are identifiable in the data. In fact, unsupervised ML can sidestep statistical assumptions entirely, for example by employing density-based methods such as DBSCAN \cite{Sonnewald2020}. Advances within ML are making it increasingly possible and convenient to take advantage of methods such as t-SNE \cite{Sonnewald2020} and UMAP, where the original topology of the data can be conserved in a low-dimensional rendition. 

Interpolation of missing data in oceanic fields is another application where ML techniques have been used, yielding products used in operational contexts. For example, Kriging is a popular technique that was successfully applied to altimetry \cite{le1998improved}, as it can account for observation from multiple satellites with different spatio-temporal sampling. In its simplest form, kriging estimates the value of an unobserved location as the linear combination of available observations. Kriging also yields the uncertainty of this estimate, which has made it popular in geostatistics. EOF-based techniques are also attracting increasing attention with the proliferation of data. For example, the DINEOF algorithm \cite{alvera2016analysis} leverages the availability of historical datasets, to fill in spatial gaps within new observations. This is done via projection onto the space spanned by dominant EOFs of the historical data. The use of advanced supervised learning, such as DL, for this problem in an oceanographic contexts is still in its infancy.  Attempts exist in the literature, including deriving a DL equivalent of DINEOF for interpolating SST \cite{barth2020dincae}.

\begin{figure}
 \centering \includegraphics[width=0.5\textwidth, angle =0 ]{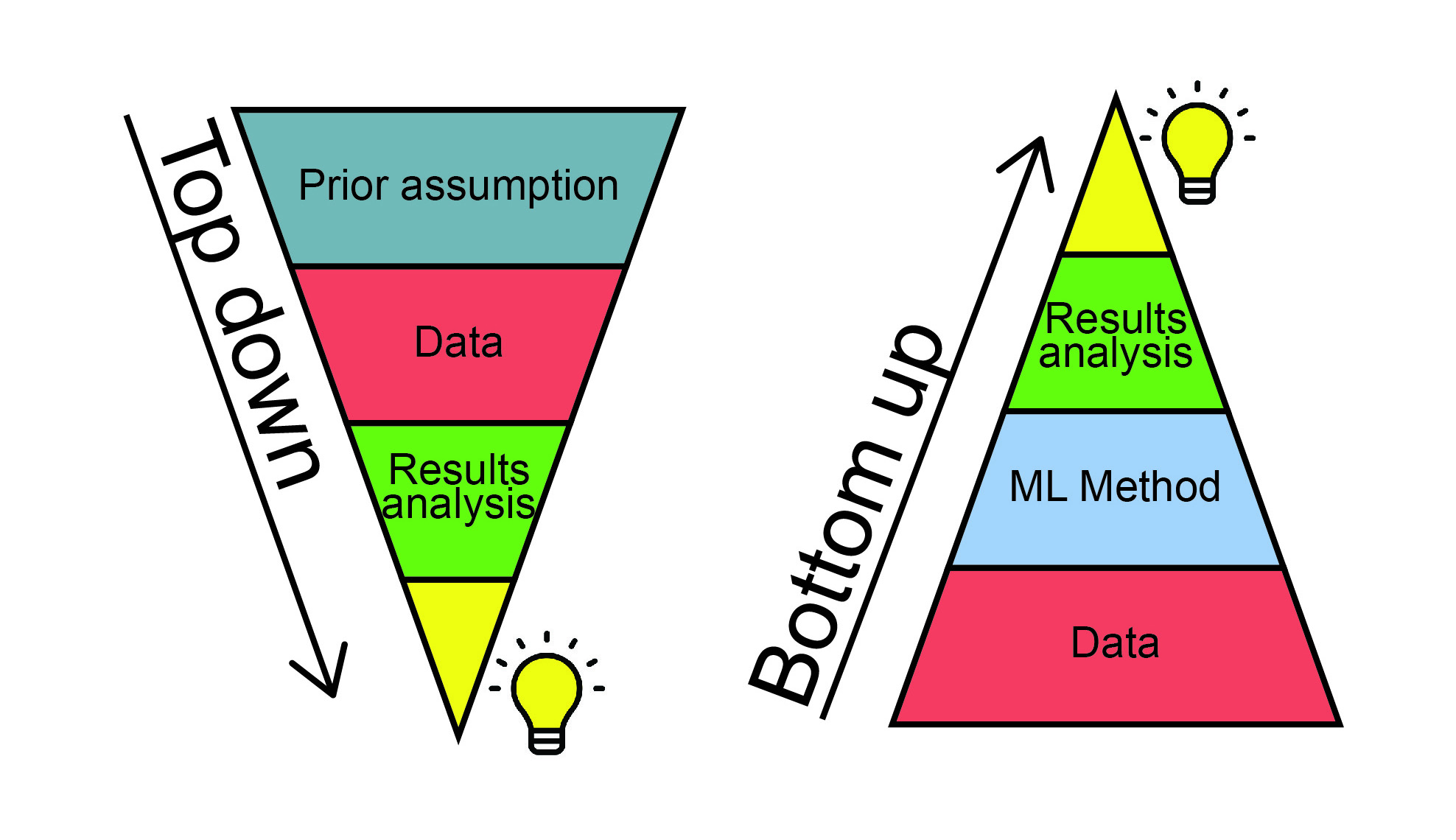}
  \caption{\textbf{Cartoon of the role of data within oceanography}. While eliminating prior assumptions within data analysis is not possible, or even desirable, ML applications can enhance the ability to perform pure data exploration. The 'top down' approach (left) refers to a more traditional approach where the exploration of the data is firmly grounded in prior knowledge and assumptions. Using ML, how data is used in oceanographic research and beyond can be changed by taking a 'bottom up' data-exploration centered approach, allowing the possibility for serendipitous discovery. }
  \label{fig:triangles}
\end{figure}

\section{Exchanges between observations and theory}
\label{sec:theory}


Progress within observations, modeling, and theory go hand in hand, and ML offers a novel method for bridging the gaps between the branches of oceanography. When describing the ocean, theoretical descriptions of circulation tend to be oversimplified, but interpreting basic physics from numerical simulations or observations alone is prohibitively difficult. Progress in theoretical work has often come from the discovery or inference of regions where terms in an equation may be negligible, allowing theoretical developments to be focused with the hope of observational verification. Indeed, progress in identifying negligible terms in fluid dynamics could be said to underpin GFD as a whole \cite{Vallis2016}. For example, Sverdrup's theory \cite{Sverdrup47} of ocean regions where the wind stress curl is balanced by the Coriolis term inspired a search for a predicted `level of no motion' within the ocean interior. 

The conceptual and numerical models that underlie modern oceanography would be less valuable if not backed by observational evidence, and similarly, findings in data from both observations and numerical models can reshape theoretical models \cite{Garabato2012}. ML algorithms are becoming heavily used to determine patterns and structures in the increasing volumes of observational and modelled data \cite{Maze2017,Jones2019,jones:19:ci,Rosso2020,Tesdal2021,sonnewald2019unsupervised,callaham2021,Houghton2020,Pauthenet2019,Boehme2021,Desbruyeres2021}. For example, ML is poised to help the research community reframe the concept of ocean fronts in ways that are tailored to specific domains instead of ways that are tied to somewhat ad-hoc and overgeneralized property definitions \citep{Chapman2020}. Broadly speaking, this area of work largely utilizes unsupervised ML and is thus well-positioned to discover underlying structures and patterns in data that can help identify negligible terms or improve a conceptual model that was previously empirical. In this sense, ML methods are well-placed to help guide and reshape established theoretical treatments, for example by highlighting overlooked features. A historical analogy can be drawn to d'Alembert's paradox from 1752 (or the hydrodynamic paradox), where the drag force is zero on a body moving with constant velocity relative to the fluid. Observations demonstrated that there \textit{should} be a drag force, but the paradox remained unsolved until Prandtl's 1904 discovery of a thin boundary layer that remained as a result of viscous forces. Discoveries like Prandtl's can be difficult, for example because the importance of small distinctions that here form the boundary layer regime can be overlooked. ML has the ability to be both objective, and also to highlight key distinctions like a boundary layer regime. ML is ideally poised to make discoveries possible through its ability to objectively analyze the increasingly large and complicated data available. Using conventional analysis tools, finding patterns inadvertently rely on subjective `standards' e.g. how the depth of the mixed layer or a Southern Ocean front is defined \cite{Dong2008, Chapman2020, Thomas2021}. Such standards leave room for bias and confusion, potentially perpetuating unhelpful narratives such as those leading to d'Alembert's paradox. 


With an exploration of a dataset that moves beyond preconceived notions comes the potential for making entirely new discoveries. It can been argued that much of the progress within physical oceanography has been rooted in generalizations of ideas put forward over 30 years ago\cite{Garabato2012,munk1951note,johnson1989size}. This foundation can be tested using data to gain insight in a ``top-down'' manner (Fig. \ref{fig:triangles}). ML presents a possible opportunity for serendipitous discovery outside of this framework, effectively using data as the foundation and achieving insight purely through its objective analysis in a ``bottom up'' fashion. This can also be achieved using conventional methods but is significantly facilitated by ML, as modern data in its often complicated, high dimensional, and voluminous form complicates objective analysis. ML, through its ability to let structures within data emerge, allows the structures to be systematically analyzed. Such structures can emerge as regions of coherent covariance (e.g. using clustering algorithms from unsupervised ML), even in the presence of highly non-linear and intricate covariance \cite{Sonnewald2020}. Such structures can then be investigated in their own right and may potentially form the basis of new theories. Such exploration is facilitated by using an ML approach in combination with IAI and XAI methods as appropriate. Unsupervised ML lends itself more readily to IAI and to many works discussed above. Objective analysis that can be understood as IAI can also be applied to explore theoretical branches of oceanography, revealing novel structures \cite{callaham2021, sonnewald2019unsupervised, Tesdal2021}. Examples where ML and theoretical exploration have been used in synergy by allowing interpretability, explainability, or both within oceanography include \cite{Sonnewald2021,Zanna2020a}, and the concepts are discussed further in section \ref{sec:synthesis}.

As an increasingly operational endeavour, physical oceanography faces pressures apart from fundamental understanding due to the increasing complexity associated with enhanced resolution or the complicated nature of data from both observations and numerical models. For advancement in the fundamental understanding of ocean physics, ML is ideally placed to break this data down to let salient features emerge that are comprehensible to the human brain. 

\subsubsection{ML and hierarchical statistical modeling}
\label{sec:hirSM}

The concept of a model hierarchy is described by \cite{ref:held2005} as a way to fill the ``gap between simulation and understanding'' of the Earth system. A hierarchy consists of a set of models spanning a range of complexities. One can potentially gain insights by examining how the system changes when moving between levels of the hierarchy, i.e. when various sources of complexity are added or subtracted, such as new physical processes, smaller-scale features, or degrees of freedom in a statistical description. The hierarchical approach can help sharpen hypotheses about the oceanographic system and inspire new insights. While perhaps conceptually simple, the practical application of a model hierarchy is non-trivial, usually requiring expert judgement and creativity. ML may provide some guidance here, for example by drawing attention to latent structures in the data. In this review, we distinguish between statistical and numerical ML models used for this purpose. For ML-mediated models, a goal could be discovering other levels in the model hierarchy from complex models \cite{ref:balaji2021}. The models discussed in Sections~\ref{sec:obs} and \ref{sec:theory} constitute largely statistical models, such as ones constructed using a k-means application, GANs, or otherwise. This section discusses the concept of hierarchical models in a statistical sense, and Section~\ref{sec:hirNM} explores the concept of numerical hierarchical models. A hierarchical statistical model can be described as a series of model descriptions of the same system from very low complexity (e.g. a simple linear regression) to arbitrarily high. In theory, any statistical model constructed with any data from the ocean could constitute a part of this hierarchy, but here we restrict our discussion to models constructed from the same or very similar data. 

The concept of exploring a hierarchy of models, either statistical or otherwise, using data could also be expressed as searching for an underlying manifold \cite{Lorenz1992}. The notion of identifying the "slow manifold" postulates that the noisy landscape of a loss function for one level of the hierarchy, conceals a smoother landscape in another level. As such, it should be plausible to identify a continuum of system descriptions. ML has the potential to assist in revealing such an underlying slow manifold, as described above. For example, Equation discovery methods shown promise as they aim to find closed form solutions to the relations within datasets representing terms in a parsimonious representation (e.g \cite{Zanna2020,ref:schmidtlipson2009,ref:gaitanetal2016} are examples in line with \cite{ref:balaji2021}). Similarly, unsupervised equation exploration could hold promise for utilizing formal ideas of hypothesis forming and testing within equation space \cite{Kaiser2021}.

In oceanographic ML applications, there are tunable parameters that are often only weakly constrained. A particular example is the total number of classes $K$ in unsupervised classification problems \citep{Maze2017,Jones2019,jones:19:ci,sonnewald2019unsupervised,Sonnewald2020}. Although one can estimate the optimal value $K^*$ for the statistical model, for example by using metrics that reward increased likelihood and penalize overfitting [e.g. the Bayesian information criteria (BIC) or the Akaike information criterion (AIC)], in practice it is rare to find a clear value of $K^*$ in oceanographic applications. Often, tests like BIC or AIC return either a range of possible $K^*$ values, or they only indicate a lower bound for $K$. This is perhaps because oceanographic data is highly correlated across many different spatial and temporal scales, making the task of separating the data into clear sub-populations a challenging one. That being said, the parameter $K$ can also be interpreted as the complexity of the statistical model. A model with a smaller value of $K$ will potentially be easier to interpret because it only captures the dominant sub-populations in the data distribution. In contrast, a model with a larger value of $K$ will likely be harder to interpret because it captures more subtle features in the data distribution. For example, when applied to Southern Ocean temperature profile data, a simple two-class profile classification model will tend to separate the profiles into those north and south of the Antarctic Circumpolar Current, which is a well-understood approximate boundary between polar and subtropical waters. By contrast, more complex models capture more structure but are harder to interpret using our current conceptual understanding of ocean structure and dynamics \citep{Jones2019}. In this way, a collection of statistical models with different values of $K$ constitutes a model hierarchy, in which one builds understanding by observing how the representation of the system changes when sources of complexity are added or subtracted \citep{ref:held2005}. Note that for the example of k-means, while a range of $K$ values may be reasonable, this does not largely refer to merely adjusting the value of $K$ and re-interpreting the result. This is because, for example, if one moves from $K$=2 to $K$=3 using k-means, there is no a priori reason to assume they would both give physically meaningful results. What is meant instead is similar to the type of hierarchical clustering that is able to identify different sub-groups and organize them into larger overarching groups according to how similar they are to one another. This is a distinct approach within ML that relies on the ability to measure a ``distance'' between data points. This rationale reinforces the view that ML can be used to build our conceptual understanding of physical systems, and does not need to be used simply as a ``black box". 
It is worth noting that the axiom that is being relied on here is that there exists an underlying system that the ML application can approximate using the available data. With incomplete and messy data, the tools available to assess the fit of a statistical model only provide an estimate of how wrong it is certain to be. To create a statistically rigorous hierarchy, not only does the overall co-variance structure/topology need to be approximated, but also the finer structures that would be found within these overarching structures. If this identification process is successful, then the structures can be grouped with accuracy as defined by statistical significance. This can pose a formidable challenge that ML in isolation cannot address; it requires guidance from domain experts. For example, within ocean ecology, \cite{Sonnewald2020} derived a hierarchical model by grouping identified clusters according to ecological similarity. In physical oceanography, \cite{Rosso2020} grouped some identified classes together into zones using established oceanographic knowledge, in a step from a more complex statistical model to a more simplified one that is easier to interpret. When performing such groupings, one has to pay attention to a balance of statistical rigour and domain knowledge. Discovering rigorous and useful hierarchical models should hypothetically be possible, as demonstrated by the self-similarity found in many natural systems including fluids, but limited and imperfect data greatly complicate the search, meaning that checking for statistical rigour is important.

As a possible future direction, assessing models using IAI and XAI and known physical relationships will likely make finding hierarchical models that are meaningful much easier. Unsupervised ML is more intuitively interpretable than supervised ML and may prove more useful for identifying such hierarchical models. Moving away from using ML in a ``black box'' sense, with IAI and XAI or otherwise, may yield a synthesis of observations and theory, allowing the field to go beyond the limitations of either; theory may allow one to generalize beyond the limits of data, and data may reveal possible structural errors in theory.

\section{From theory to numerical models}
\label{sec:models}

The observation of patterns in data is the precursor to a theory, which can lead to predictive models, provided theory can be converted to practical computation. In this section, we discuss how ML could change the way theory is represented within ocean modeling. To represent the ocean using numerical models is to help filling in missing information between observations. In addition, models act as virtual laboratories in which we can work to understand physical relationships. For example, how the separation of boundary currents such as the Gulf Stream depends on local topography or boundary conditions. The focus of this discussion will be on models that represent the three-dimensional ocean circulation, but most of these ideas can also be used in the context of modeling sea-ice, tides, waves, or biogeochemistry. We also discuss a recurring issue within ocean modeling: the presence of coastlines that complicate the application of methods that are convolutional or spectral.  


\subsection{Timescales and space scales}
\label{sec:scales}


When building numerical models, the ocean is largely treated like a typical fluid that follows the Navier-Stokes equations, and the challenges faced therein are similar to those presented by general computational fluid dynamics. The filamentation of the flow results in scale interactions that make it necessary to represent all spatial scales within the model, while the model resolution needs to be truncated due to the finite nature of computational power. The dynamics at different scales can either be represented via the explicit, resolved representation within the model or via the parametrization of sub-grid-scales as a turbulent closure. 

Much research has gone into the formulation of parametrizations to represent the sub-grid-scales. Such representations range from classical closures for turbulent fluids, using formulations such as Gent-McWilliams \cite{Gent1990} that take the dynamics of sub-grid ocean eddies into account, to empirical closure schemes that are determined by comparing simulations at a target resolution to simulations at higher resolution \cite{Cooper2015,Ryzhov2020}. Lately, ML has also been used to learn the sub-grid-scale, either via the direct learning of the terms using NNs \cite{Bolton2019} or via the learning of the underlying equations \cite{Zanna2020}. Similar and promising DA applications are also emerging, discussed in section \ref{sec:DA}.

Next to the representation of the sub-grid-scale, numerical ocean models are also prone to errors due to the necessary discretization of the differential equations on a numerical grid. A number of methods are used to discretize the equations \cite{Durran2010}, including finite difference, finite volume, and finite element methods. In comparison to the atmosphere, spectral discretization methods cannot easily be applied to the ocean due to the presence of coastlines, as creating a representation using global basis functions is not straightforward.

In the presence of perfect data and adequate computational power with which to train a DL application, it would be theoretically possible to learn the dynamics of the ocean with no knowledge of the equations of motion. This is because DL can learn the update of the physical fields based on time-series of observations or model data. This has been done successfully for certain atmospheric applications \cite{Dueben2018,Weyn2020,Rasp2020} and for an idealized ocean model \cite{Furner2021}. However, DL representations of the ocean are more difficult than for the atmosphere. This is because there is much less reliable three-dimensional training data available for the ocean spatially, and because relevant time-scales of the ocean are much longer together with the shorter time scales that together make up the ocean state. This is because the ocean has important low-frequency variability, resulting in a need for longer training data sets. Furthermore, coastlines form lateral boundaries that may reduce the quality of NN solutions. This is because, NN models often require a certain stencil of local information to update the physical fields at a given gridpoint. For example, CNNs perform best if the underlying system is invariant by translation. While grid-points on land could be incorporated into local stencils with predefined values, the presence of coastlines may reduce the amount of training data for specific pattern of the coast line. Also, having sharp discontinuities from ocean to land pixels results in a more challenging problem for NN in general. For example, a CNN could spend a considerable amount of its parameters learning the coastline boundary patterns, which may not be of interest if the user is focusing on ocean-only patterns.  

ML tools could also serve as a method to represent the ocean with fewer degrees of freedom than a full conventional numerical model. Such use cases for ML include being used  (1) as part of a coupled Earth system model that is either used for short-term weather forecasts, or (2) in long climate simulations. For example, if a model is only trying to represent the surface fields that are most important for the coupling to the atmosphere, the model could focus on the use of the leading principal components (if these can be derived in the presence of coastlines), and learn the interactions between the different components using data from a time-series extracted from long model (or observational) trajectories. Here, a first approach towards building low-order ML models using a barotropic model showed that results from high-dimensional ML tools from DL, such as complex RNNs, may not necessarily provide better results when compared to ML techniques that are based on regression techniques and stochastic forcing \cite{Agarwal2021}.

The surface of the ocean has strong and direct impact on society, and ML tools may help to better understand and characterize its dynamics. For coupled models, the importance of representing the interactions between the atmosphere and sea ice components is well established. However, allowing this interaction involves major computational challenges and uncertainties, associated with understanding of the complex feedback and coupling processes between system components. Many surface processes, for example ocean waves and storm surges, are typically excluded from long-term ocean simulations and modelled separately, although exceptions exist \cite{Song2020}. This exclusion is largely because of the very large differences in phase speeds, for example between surface waves and the deep ocean that result in a barotropic (fast) component being treated separately from a baroclinic (slow) component \cite{Tolman1991,Griffies2001}. However, there are now approaches to improve wave and sea-ice modelling or predictions using ML, that either try to improve on the computational efficiency or the accuracy of conventional methods \cite{Puscasu2014,Andersson2021,Zhai2020, Choi2019, Palerme2021}. 



\subsection{Concepts of ML and hierarchical numerical modeling}
\label{sec:hirNM}

This section discusses hierarchical modeling in a numerical sense, complementing Section~\ref{sec:hirSM} that discusses hierarchical modeling in a statistical sense. Within oceanography, observations and theory are more meaningful when viewed together. Observational scientists (see Section~\ref{sec:obs}) make choices of what to sample based on some prior conceptions of relevance, and of course theory is ungrounded without data. In epistemology, this is often summarized in Duhem's formulation, ``theory is data-laden, and data is theory-laden'' \cite{ref:duhem1906}. In talking about climate and weather modeling, Edwards made the corollary, ``models are data-laden, and data is model-laden'' \cite{ref:edwards2010}. For example, the concept of a reanalysis dataset comes from a model. The sequence from observations to theory to models and predictions shows this interplay. This is a key sequence where we expect ML to display its strengths, e.g. where IAI and XAI methods may yield a synthesis of observations and theory, allowing one to go beyond the limitations of either: theory allowing one to generalize beyond the limits of data, and data revealing possible structural errors in theory as detailed in Section~\ref{sec:theory}. Ideally, we would like to go beyond these and use ML to discover the underlying equations (e.g \cite{ref:bruntonetal2016}), and deliver a model hierarchy that can then be implemented numerically (\cite{ref:held2005},\cite{ref:balaji2021}). While simple in principle, in practice this concept is less straightforward to implement. An example of a form of equation discovery can be seen in Zanna and Bolton's \cite{Zanna2020} reduction of resolved turbulent dynamics into a representation suitable for use in coarse-grained models. The coarse-grained models represent a different level of the hierarchy, if tiers are set by horizontal resolution. This ML model was arrived at applying an RVM, with the different equation terms serving as the input. This is an example where the results of an ML application yield a parsimonious closed-form representation, and are therefore interpretable. Using XAI, it would also be possible to infer what gave the ML application its predictive skill, which could eliminate e.g. contamination from numerical issues that are model resolution specific. Methods constituting equation discovery are an exciting, and potentially powerful, way ML could impact numerical modeling, particularly if IAI/XAI can be applied to ensure the ML application predictive skill is grounded in physics.


\subsection{Computational challenges}
\label{sec:compute}

Since the first ocean general circulation model \cite{bryan1968nonlinear,bryan1997numerical}, available computational power has grown exponentially, following Moore's law. The realization that the ocean is fundamentally turbulent and topographically influenced \cite{Wunsch2002, Garabato2012} resulted in numerical model development focused on increasing model complexity and refining the model discretization. Numerical model performance is often measured in simulated years per day (SYPD). Computational challenges largely manifest as a balance between preserving the significant legacy present in current ocean modeling codes and harnessing the significant advances within the field of high performance computing, which is often tailored to ML. ML is a trillion dollar industry which is based on high-performance computing power \cite{McKinsey}. It is therefore driving developments in modern supercomputing.


The growth of processing speed in supercomputers is no longer exponential, but improvements in the computational efficiency of ocean models are still possible through customisation of the computing hardware. ML may likely have a place within a revision of ocean models to improve their computational efficiency. Even within Earth system models as a whole, a ``digital revolution" has been called for \cite{Bauer2021}, where harnessing efficiency in modern hardware is central. Computers can increasingly be customised as hardware is becoming more heterogeneous, meaning that different components for data movement and processing can be combined \cite{Bauer2020}. Examples of such heterogeneous hardware include the so-called Graphical Processing Units (GPU), Tensor Processing Units, Field-Programmable Gate Arrays, and Application Specific Integrates Circuits, which largely are highly compatible with ML. To take advantage of this heterogeneous hardware, making current ocean models ``portable'', a significant effort would be necessary \cite{Bauer2020}. Current ocean models use the Fortran programming language and are parallelised to run on many processors via interfaces such as MPI and OpenMP. This parallelisation approach is not compatible with hardware accelerators such as GPUs. Compatibility could be achieved via re-writing or enhancement by programming interfaces such as OpenACC or Cuda. Some model groups are investigating a move to newer computing languages, such as Julia (such as the Oceananigans model as part of the CliMA project \cite{Ramadhan2020}). Languages like Julia can hide technical details in high-level descriptions of the model code making it more portable. So-called domain-specific languages can be used to facilitate portability \cite{Gysi2015}. Here, the main algorithmic motives are formulated into library functions that can be ported to different systems with no need to change the model code that is used by the model developer.

ML is expected to play a role in issues associated with the purely computational approach to ocean modeling, beyond devising portability to different hardware accelerators such as GPUs. Hardware accelerators are best suited to problems of high operational intensity (floating-point operations per memory operation). The discretized differential equations governing fluid flow typically result in sparse operations resulting from near-neighbour dependencies (``stencils''). Stencil codes remain notorious for their low operational intensity \cite{ref:barbayokota2013} resulting in poor computational performance, and despite substantial efforts in recent years there has been little progress \cite{ref:neumannetal2019, ref:balajietal2017}.
This problem is accentuated in oceans, whose long timescales often require \order{(1000 SYPD)} for the basic dynamics to emerge. The role of ML in emulating turbulent ocean dynamics is likely to be critical in achieving the level of performance required. This is because resolving key phenomena such as mesoscale eddies remain computationally out of reach, and the current parameterizations such as from Gent and Mc Willians (1990) \cite{Gent1990} discussed in Section~\ref{sec:scales} continue to exhibit deficiencies in simulating meridional eddy transport \cite{ref:fox-kemperetal2014}.

ML, and in particular DL, could play a significant role in improve computational efficiency of ocean models due to its ability to work with low numerical precision. Many operations are memory bandwidth bound, and as DL is based on dense linear algebra it is capable of working with very low numerical precision, such as IEEE half precision with 16 bits per variable \cite{Kurth2018}. 
The trend towards ML hardware that is optimised for dense linear algebra and low numerical precision may have an impact on future ocean modeling. The use of low numerical precision has been discussed for weather and climate models \cite{Dueben2014}. The NEMO model~\cite{madec_gurvan_2019_3878122} was run in single precision with 32 bits per variable instead of the default of double precision with 64 bits per variable \cite{Tinto2019}, and half precision with 16 bits per variable is being explored for weather and climate models \cite{Kloewer2020} and hardware that is customised for ML has been tested to speed-up expensive components of conventional models \cite{Hatfield2019}. However, in particular for the long-term simulations needed in the ocean, care needs to be taken to make sure that rounding errors do not impact on conservation laws. Certain specific aspects of ocean dynamics require a large dynamic range. For instance, sea level rise, which is a secular change measured in cm/century, must be simulated against a backdrop where surface waves have an amplitude measured in $\mathcal{O}(m)$ and a phase speed of $\mathcal{O}(100 m/s)$, at least 8 orders of magnitude larger over a typical ocean timestep. For subsequent analysis, it is worth noting that using lower numerical precision would also impact the ability of doing analysis on budgets, as closing these can be complicated when rounding errors are biased.

ML is being explored as a method to emulate computationally costly components of ocean models. This was done successfully in a number of studies \cite{rasp2018deep,brenowitz2018prognostic} for physical parametrisation schemes of atmospheric models. For ocean modeling, NN emulators could for example speed-up biogeochemical components \cite{Nowack2018}, which often form a large cost-fraction for ocean models in climate predictions, or sea-ice models, which are often a computational bottleneck as they are difficult to parallelise. ML could also be useful for improving advection schemes and learning local corrections and limiters of fluxes between grid-cells \cite{Kochkov2021}. Furthermore, it may also be possible to improve efficiency of ocean models with semi-implicit timestepping schemes. Here, ML could be used to precondition solver for the large linear problem that needs to be solved in every timestep by estimating the results \cite{Ackmann2020}. 


The exponential growth of computing power has been accompanied by an exponential growth in data volume. This growth represents a big challenge for operational weather and climate predictions \cite{Balaji2018}. As data movement is very expensive and a bottleneck in performance, ocean models need to be ``data-centric" and the workflow of the model should be designed in a way that would reduce data movement to a minimum. For example, data is conventionally simply written to discs or tapes after a model simulation, to be retrieved by users afterwards for analysis. A data-centric workflow would process data on-the-fly before it is stored. ML, and in particular unsupervised ML, would be essential in enabling domain scientists to extract the relevant information in such a data-centric workflow. However, such a workflow would also results in additional requirements in terms of the training of staff and the software and hardware infrastructure of weather and climate centres \cite{Dueben2021}. For example, it is not trivial to incorporate ML tools that are commonly developed using the programming language Python, into atmosphere or ocean models that are typically written in Fortran. Efforts towards such integration are taking shape, with libraries in development that link NNs into Fortran code \cite{Curcic2019,Ott2020,partee2021using} 

Also of note is the increased difficulty in extracting scientifically interesting information from the vast amounts of data produced by numerical models that \textit{is} stored. The complexity and sheer size of many of these data hinder data dissemination and analysis and severely hamper efforts to analyze the data and address research goals. This emerging class of problems can be illustrated by the Coupled Model Intercomparison Project (CMIP) ensemble now in its sixth phase, which is expected to generate an estimated 40,000 TB of climate model data, a 20-fold increase in data volume from the previous phase \cite{eyring_towards_2016,CMIP6}. Many variables needed for analysis are effectively unavailable due the difficulty in saving or sharing the data. ML has the capacity to efficiently analyze large datasets as shown in section \ref{sec:obs} and \ref{sec:theory}, but it has also been used to infer, for example, information about sub-surface currents \cite{chapman2017reconstruction,Manucharyan2021}, eddy heat fluxes \cite{George2019} and full 3-dimensional dynamics in CMIP6 \cite{Sonnewald2021}. ML in many forms has the potential to be highly valuable for researchers interested in the analysis of data that is increasingly large, potentially sparse, and partially unavailable for logistical reasons \cite{Eyring2019}.

\subsection{Enforcing physical priors in ML methods}

When physical constraints are enforced within ML techniques, this is equivalent to incorporating physical understanding into the applications. Using statistical language, we can describe this process as ``enforcing physical priors''. ML techniques backed by massive datasets have achieved groundbreaking results in vision, speech, and natural language processing, but they have yet to reach the physical oceanography community or largely the physical sciences in general. The ocean is governed by complex phenomena that have been studied by oceanographers for centuries, and taking advantage of this scientific heritage is one way of helping ML techniques reduce the search space of solutions, i.e. by guiding them using physical theories. This research direction is increasingly attracting attention as it helps constrain ML algorithms to be physically plausible and facilitates the interpretation of the results by domain experts. There is a broad spectrum of techniques to supplement ML with physical constraints \cite{willard_integrating_2020}, of which only the most directly relevant are discussed here.

The simplest way to enforce physical priors is through the loss function used to train the ML model. Concretely, this is done by adding an error term related to the physical constraint that needs to be respected, such as a conservation law. For example, if the output field $F$ in a regression problem need to be divergence-free, the term $\Vert \nabla F\Vert$ is added to the total loss function to ensure that the divergence of $F$ is close to zero. This approach has its mathematical roots in the theory of Lagrange multipliers. It can also be seen as a way of doing \textit{regularization}, meaning that finding solutions that generalize well to unseen data is more likely. However, adding physical priors as terms in loss functions comes with a price, which is the problem of weighting different loss terms to impose which ones are most important. The problem of weighting can be solved using cross-validation techniques. With cross-validation, a holdout dataset called a validation dataset is left apart, and the weights of the losses are tuned to achieve the best performance by comparison to this validation dataset. However, such cross-validation techniques can be difficult when the number of constraints is high.

A second strategy that has gained much attention in the recent years is enforcing the constraints directly in the mapping function used for learning. This strategy is best suited to NNs given their flexibility and the rich design choices that enable them to be tailored to specific data. The NN architecture is designed with the physical priors in mind. For example, if we already know that the quantity we want to find is a multiplication of two quantities, then we can encode this inside the neural net by creating two sub-networks whose outputs are multiplied in the last layer~\cite{fablet2017bilinear, brajard2020combining}. 

While enforcing physical priors has been a very active area of research in the atmospheric community (see section~\ref{sec:mlESS}), few papers investigating the potential of combining ML and physics can be found in the ocean science literature. In the following we cite some of these examples. Authors in \cite{bolton2019applications} reconstruct subgrid eddy momentum forcing using ConvNets and found that enforcing a constraint on global momentum conservation can best be done by either postprocessing the ConvNet's output or hardcoding a last layer in the ConvNet that removes the spatial mean of the data. \cite{Zanna2020a} proposes to use an equation discovery algorithm, namely Relevance Vector Machines (RVM), for ocean eddy parameterizations. Few attempts have been made to forecast ocean variables using a mix of physical models and DL tools, notably in \cite{bezenac_deep_2019} where authors model an advection diffusion equation in a DL architecture used to forecast SST, while \cite{erichson2019physics} tackle the same problem by combining an autoencoder with ideas from Lyapunov analysis, and \cite{lguensat2019learning}, where a NN is embedded inside a one-layer quasi-geostrophic numerical model to reduce its bias towards a 3D ocean model.

Enforcing physical priors by solving differential equations with ML techniques is an active research direction that features the development of interesting tools for the ocean community, which are still under-exploited. Physically Informed Neural Networks (PINNs) \cite{raissi2017physics} is a notable example of a technique that leverages the power of NNs to solve differential equations such as the incompressible Navier-Stokes equation \cite{jin2021nsfnets} without a need for mesh generation, which could accelerate model development. Other recent techniques for learning ordinary differential equations using either NNs \cite{chen2018neural} or a combination of NNs and physical-based components \cite{rackauckas_universal_2020} are a promising line of research at the interface of NNs and differential equations, which to the best of the authors' knowledge has not yet been applied to ocean modeling.


\section{From models to predictions}
\label{sec:predict}

A basic goal and a test of the understanding of a physical process is the ability to predict its behaviour. Predictions of the weather for several days are a major geoscientific success. Such forecasts have improved with the increasing availability of computational power and observational networks, as well as better algorithms and process understanding \cite{ref:baueretal2015}. However, predictions of the Earth system on longer timescales are still a major challenge. This is problematic, as predictions often form the basis of decision making. Understanding model error, and combining models with observations, is also at the core of supporting decision makers discussed in Section~\ref{sec:synthesis}.

\subsection{Model bias and model error}
\label{sec:bias}

Bias and error in models are addressed through a systematic process of improvements in our understanding, but the needs of decision support can be immediate. Constraining simulations using observations is the process of data assimilation, covered below in Section~\ref{sec:DA}. But where errors are recalcitrant, oceanographers and applied scientists in general use methods of ``artificial'' error reduction, driven by comparisons against data. An early example of correcting for a bias related to the ocean's role in climate is the use of ``q-flux adjustments',' or simply flux adjustments. Here, the issue was a persistent error in the evaporative flux from the ocean surface. The adjustment to ameliorate this bias was a correction to restore energy balance to the coupled system by artificially adding a compensation term\cite{ref:manabeetal1991}. This adjustment method fell into disfavor owing to its blatant ``fudge factor'' nature \cite{ref:shackleyetal1999}, although recent studies indicate that ``flux adjusted'' models continue to exhibit greater predictive skill \cite{ref:vecchietal2014}.

When assessing a prediction from a model, the accuracy of the output can be assessed by comparing to a 'truth' benchmark. Such a benchmark can for example be from observations or a target model representation of the system. Observations, although mostly not complete, constitute a best guess. This process can also identify ``structural error'', also mentioned in Section~\ref{sec:mlESS}, indicating that the model formulation itself is incorrect. Compared to observations, model outputs can show differences that cannot only be attributed to differences in initial conditions, but instead reflect errors within the model itself discussed in Section \ref{sec:DA} below. Some of these errors can be explained by unresolved scales in the discretized version of numerical models, but model errors can also originate from incomplete physical knowledge. For example, within a sub-gridscale parameterisation the exact physics that need to be represented may be unclear, as discussed in section \ref{sec:models}. Incomplete physical knowledge also impacts uncertainties in the parameters used, for example in the coupling terms between model components. Within model error as a whole, there may be a systematic component, which is referred to as model bias.

For post-processing of model output, statistical methods, related to ML, have been used to correct biases (for example \cite{Schuhen2012,Laloyaux2020} or flux adjustments). Bias correction methods are used frequently in operational weather predictions with DL playing an ever-increasing role \cite{Rasp2018b,Baran2020,Groenquist2020}. However, using downscaling as described in Section~\ref{sec:mlESS}, ML can also be used to relate model output with local information, such as the local topography at very high resolution or observations that are available, to improve predictions when model simulations have already finished. Called up-scaling within the ML community, some of the mapping procedures used for downscaling, such as GANs, even allow for uncertainty quantification \cite{Leinonen2020}. Within climate models, the LRP XAI method have successfully been used to identify key model biases for certain prediction tasks \cite{barnes2020identifying}, with potential for application to the ocean. However, the LRP method application is still in its infancy. 




\subsection{Ocean data assimilation}
\label{sec:DA}

\subsubsection{Data assimilation methods: A brief history and main assumptions}
\label{susec:DAintro}

Data assimilation (DA) is the process of constraining a theoretical representation of a system, usually using a numerical or statistical model, using a collection of observations. The results of this process typically include optimized estimates of (1) the time-evolving state of the system (sometimes called the ``trajectory"), (2) initial conditions, (3) boundary conditions, and (4) other intrinsic model parameters (e.g. mixing coefficients). The optimization process typically consists of correcting the values of the initial conditions, boundary conditions, and model parameters in order to minimize a selected model-data misfit metric. To use the language of the theory of differential equations, one may think of DA as a set of methods for rigorously identifying which solution, among the family of solutions to a system of differential equations, best satisfies the given constraints. 

Although there is a long history of DA in numerical weather prediction stretching across much of the 20th century, oceanographic DA only began in the late 1980s. The first experiments were regional~\cite{robinson1989data}, followed a few years later by the ambitious World Ocean Circulation Experiment (WOCE, \cite{Wunsch2005}), and a community was subsequently assembled under the Global Ocean Data Assimilation Experiment (GODAE, \cite{godae}). These first DA approaches used in weather and ocean prediction were directly derived using optimal interpolation~\cite{ghil1991data} and were based on strong assumptions, namely that the evolution model is linear and perfect and that the data error distribution is unbiased and well-represented by a Gaussian. In time, DA algorithms evolved to relax some of these assumptions, extending the scope of DA applications to the ocean. 

The developments within DA have led to two main sets of techniques. These are ensemble approaches, of which the ensemble Kalman filter (EnKF) is a standard example, and variational approaches such as four dimensional variational assimilation (4DVar). Both classes of methods conceptually represent the abstract trajectory of the target system as a probability distribution across possible trajectories. EnKF constructs an ensemble of forecast states such that the ensemble mean and the sample covariance are expected to be the best estimates. A core assumption is that the ensemble probability distribution can be well-represented by a Gaussian function \citep{evensen1994sequential}. The 4DVar method uses a linear model to calculate which perturbations to the initial conditions, boundary conditions, and parameters tend to increase the agreement between the time-evolving state of the model and the observational constraints \citep{courtier1994strategy}.

Each of the DA classes of methods are used in their various flavours for both global or regional studies \citep{mazloff_eddy-permitting_2010,forget_ecco_2015,verdy_data_2017,aste_2020, lellouche2018mercator, sakov2012topaz4}. DA is used routinely both in operational forecast and reanalysis mode. DA is used in the framework of several national and international projects. In no particular order, examples include the  ECCO\footnote{Estimating the Circulation and Climate of the Ocean} project, ECMWF\footnote{The European Centre for Medium-Range Weather Forecasts} or the NOAA NCEP\footnote{National Oceanic and Atmospheric Administration, National Centers for Environmental Prediction} Global Ocean Data Assimilation System (GODAS) in the USA.

In idealized comparisons between the two classes of methods, EnKF produces more accurate estimates for shorter assimilation windows, whereas 4DVar produces more accurate estimates when data constraints are sparse. For ocean applications, data is often sparse, making 4DVar attractive \citep{kalnay:2007}. In practice, different DA approaches derived from optimal interpolation, 3DVar, the EnKF, or 4DVar are used~\cite{cummings2009ocean}. The type depends on the application (e.g. short-term forecast or climate application), the available computing resources, the type of observations that are assimilated, and the historical expertise in each group.

\subsubsection{Model errors and ML within data assimilation}
Historically, DA techniques mainly focus on the estimation of the state of the system, but the estimation of model error in the DA process is increasingly important ~\cite{carrassi2010accounting}. Several approaches that are used to handle model error apply DA frameworks that can be considered ML approaches~\cite{Tandeo2020review, cocucci2021model}. The estimation of model errors is particularly important if DA is being used to calculate forecasts over long timescales, i.e. from sub-seasonal to decadal scales. This is of particular importance for ocean forecasts, where timescales are longer than in the atmosphere; DA has been shown to be effective in this context~\cite{wang_towards_2020}.
 
\subsubsection{Data assimilation and ML}
Several studies have highlighted the connection between DA and ML~\cite{abarbanel2018machine, bocquet2019data, brajard2019connections, geer2021learning}. The connection is more direct with 4DVar, in which a function that quantifies model-data disagreement (i.e. a ``cost function'') is minimized using a gradient descent algorithm, wherein an adjoint model calculates the gradient. In this perspective, 4DVar is approximately equivalent to the process of training of a neural network for regression. This is because the adjoint model can be seen as equivalent to the gradient backpropagation process~\cite{hsieh1998applying, ref:kovachkistuart2019}.

There are several ways ML can be used in combination with a DA framework. First, a data-driven model can be used to emulate a numerical model, partially or totally to provide the forecast. The objective is then to correct the model error, or to decrease the computational cost \citep{lguensat2017analog}. Note that emulation could become instrumental, since DA methods increasingly rely on ensemble runs, which are costly~\cite{carrassi2018data}. As DA allows one to bring the model and observations close enough together to represent the same physical situation, DA can in principle be used to extend the learning of parametrization to the learning of improved models directly from observations~\cite{brajard2021hybrid, bonavita2020machine}, described further in the section~\ref{sec:models}. It is still unclear whether observations are too sparse for this approach to be successful within ocean modeling. This is particularly the case, because the time period where dense observations are available is relatively short, compared to the long timescales that are known to be important for ocean dynamics. Another benefit of using an ML emulators arises because most ML tools, such as NNs, are easy to differentiate. Given the structure of NN (interconnected simple operators), and the libraries used to implement them, the computation of the gradient of the NN model is straightforward and efficient. This means that the computation can be used to efficiently develop tangent linear and adjoint model code, which is required for DA methods such as 4DVar \cite{Hatfield2021}. This is noteworthy, because traditionally the development of tangent linear and adjoint models has required major efforts from the research community, either by manually coding an adjoint or by the semi-automatic process of algorithmic differentiation (e.g. \cite{giering_recipes_1998}). 

Second, ML can be instrumental in strongly coupled DA. Strongly coupled DA consists of correcting a coupled system (e.g. ocean-atmosphere) in a unified way. This allows, for example, atmospheric observations to constrain the ocean state and vice versa, which is not the case in uncoupled DA, where only ocean observations are used to constrain the ocean system. Strongly coupled DA is expected to be efficient but challenging due to the high variety of temporal and spatial scales~\cite{penny2019strongly}. In this sense, ML can be used to relax some strong assumptions of the DA algorithm (e.g. the assumption that the errors follow a Gaussian distribution), or to isolate relevant scales in observational and model states. Traditional DA algorithms correct the state of the model either directly, or by using a linear transformation. ML could extend this approach by applying a non-linear transformation to the space before computing the DA correction. Some examples of this approach have been developed~\cite{amendola2020data, mack2020attention, grooms2021analog, fablet2020learning}, but so far none of them have been applied to realistic ocean DA setups.

Finally, ML can help deal with the mass of available observations. In section~\ref{sec:obs}, we discussed how ML can help derive new type of products from observations. These new products are good candidates for inclusion in a DA system. ML can also be used to provide more accurate and/or faster observation operators, for example to emulate satellite observations~\cite{bonavita2020machine}.

\section{Discussion: towards a new synthesis of observations, theory, modeling, and prediction in ocean sciences using ML}
\label{sec:synthesis}

\subsection{The need for transparent ML}
\label{sec:transpML}
To increase confidence in the use of ML, stepping out of the ``black box'' is advisable. Towards this, having ML methods be transparent is very important. A transparent ML application is one where source of skill is known, or put differently why the ML came to its conclusion.  Possibly the largest hurdles for ML adoption are a lack of trust and the difficulty of generalization. These two are linked, and if generalization is not reached trust is certainly not merited for ML applications within oceanography. Generalization refers to a model's ability to properly adapt to new, previously unseen data. Within oceanography and beyond, the ideal generalization would come from the ML application learning the underlying physics. With a lack of good data coverage, the possibe underspecification \cite{damour2020underspecification} and shortcut learning \cite{Geirhos_2020} are important to keep in mind, where a model can seemingly perform well for example in the current climate but will fail in a future scenario as something physical was not learned. These issues are ubiquitous and not unique to oceanography or to Earth science, with a call for `physics informed' ML\cite{reichstein_deep_2019}. Accordingly, the field of ML already has, and is, developing methods to address these issues such as ``few shot learning'' and ``transfer learning''. Here, for ``few shot learning'' the ML models are developed to deal with very small amounts of training data \cite{wang2020}. Using ``transfer learning'' refers to where NN are trained on general tasks, but are used and adapted to specific applications without the need to retrain from scratch. If it is possible to reliably quantify and account for the uncertainties associated with an ML application during training, this could also increase confidence in the model, but assessing the reliability may face challenges similar to underspecification. Recent works show promising progress toward this by learning a probability distribution of outcomes that can be stochastically sampled\cite{ref:dunbaretal2020},\cite{ref:guillauminzanna2021}. For uncertainty quantification, the uncertainty could be determined during training with ML, likely increasing the reliability of the results. Other methods such as regularisation, invariances, dimensionality reductions are also a powerful tool to increase the generalization skill. For climate applications, a key issue when training ML applications is that the system they are being trained on is largely non-stationary. This complicates the problem of generalization even further, but ML methods have demonstrated that having good generalization skills in a non-stationary context is possible~\cite{Patel2021}. Increasingly, the ML community is suggesting a focus on using IAI \cite{Rudin2019}, driven among other things by the consistent racial and gender bias revealed in DL applications. With the ability to interpret the ML model itself, and intuitively discern if it is meaningful, the danger of introducing such bias is likely reduced dramatically. Similarly, XAI methods for example for NN, that retrospectively explain the source of ML predictive skill, can also help inspire confidence \cite{mcgovern_making_2019,toms_physically_2020,Ebert-Uphoff2020}. XAI methods such as layerwise relevance propagation (LRP \cite{Olden2004,Bach2015}) have been gaining traction within the atmosphere\cite{barnes2020, barnes2020identifying, brenowitz2020, toms_physically_2020}, and ocean, but making their application explicitly appropriate to oceanography, and indeed the physical sciences in general, may require targeted method development. 

\subsection{Decision support}
\label{sec:decsup}


There is a need for accurate and actionable information about the ocean for a wide range of decision making. As noted above in Section~\ref{sec:bias}, the need for actionable predictions and decision support can short-circuit the scientific process of error elimination. This is because the information may have ``customers''/users with an immediate need: for example decisions on shorelines ranging from building seawalls, issuing housing permits, to setting insurance premiums. ML may play a role in bridging the gap between what model-based predictions are able to provide, and what users wish to know. The role of data-driven methods could be particularly important for filling in the gaps where theory and models underspecify the system, potentially leaving considerable uncertainty as noted in Section~\ref{sec:DA}.

The reliable quantification of uncertainties is often essential to support decision making. However, uncertainty quantification is often difficult for conventional approaches used in ocean science. This is because model errors cannot be described by physical equations or physical reasoning in most applications, and errors are often noisy and non-linear. On the other hand, model error can often be diagnosed against a reference truth, such as observations or target model simulations. Therefore, ML can be useful for the quantification of uncertainties. In particular as datasets from different data sources, on different reference grids and for different variables can be fused and compared using ML techniques. For example, ML can be used to post-process ensemble simulations \cite{Schuhen2012}, and Bayesian ML techniques can also be used to learn the uncertainty quantification together with the ML tool. In addition, targeted loss function design could help target, and lessen areas of uncertainty important for specific decisions. In an ML context, the models can be calibrated or tuned toward a particular loss function \cite{ref:hourdinetal2017}. A loss function could be designed to capture physical constraints, such as the closure of the energy budget in climate simulations. However, loss functions can also be specifically designed for use cases in decision support. An open area of research remains in relating the results that may be obtained from different calibrations (loss functions) of the same system trained on the same data.

ML can help to map model data and observations to predict or detect events for which we cannot provide a useful physical representation of the interactions. This could, for example, be a mapping from observational data of a time series in a specific location or observations from a buoy, to large-scale model data with the goal of making customised predictions of surface waves and local wind. Such data could for example be used for a sailing competition. Such tools based on ML could become essential for decision support, for example when used to predict sea levels \cite{Zust2020}. ML based mapping tools could also be useful to inform where more observational data is needed, for example when deciding where to sample on a cruise or where to send autonomous platforms. To date, satellite images are largely used, but added guidance from ML techniques could be very valuable, particularly if sub-surface observations are the target \cite{Manucharyan2021, chapman_reconstruction_2017, Sonnewald2021}. ML may eventually be used to support observational campaigns in near-real time by interactively connecting networks of non-autonomous and autonomous observing platforms (e.g. gliders) to decision planning systems. These systems can take environmental conditions, target observations, and task scheduling into account. The vision is to have a ``cyberinfrastructure" that can maximize the spatiotemporal coverage of the observations without a specific need for human intervention. The potential use of such observational planning and adjustment systems is being explored by international initiatives such as the Southern Ocean Observing System (SOOS, \url{https://www.soos.aq/}). Similarly, for planing legislation, having knowledge of what is within a nations marine area and how this may connect to the surrounding ocean can be very valuable. Here ML has been used to provide actionable information~\cite{Sonnewald2020}, as the ocean does not adhere to borders drawn by humans.

Next to DL methods, the calibration of parameters is very important as many within atmosphere and ocean models cannot be validated within their physical uncertainty range. Because they cannot be validated, the parameters need to be tuned \cite{Tuppi2020,cleary2021}. Given this physical uncertainty, using ML and DL in particular will likely be very valuable as noted in Section~\ref{sec:models}. If successful, such breakthroughs could help inform a wide range of decisions including those based on climate models such as CMIP, or in a more general sense. This is particularly the case for longer timescale integrations from seasonal and onward, due to the longer timescale active within the ocean.


An important component of supporting decision makers is communication. The ability to communicate effectively between the people that are making decisions and oceanographers can pose a problem. Oceanographers would need to be aware of what \textit{is} useful information, and how to provide this. Decision makers largely may not have intimate knowledge of what available tools are capable of addressing, but mainly knowledge of the problem at hand. While seeming trivial, improving this line of communication is an important component of increasing the utility of oceanographic work.

\subsection{Challenges and opportunities}
\label{sec:chalopp}


In this review, we have highlighted some of the many challenges within observational, computational, and theoretical oceanography where ML offers an exciting opportunity to improve the speed and efficiency of conventional work and also to explore completely new avenues. As a merger of two distinct fields, there is ample opportunity to incorporate powerful, established ML methods that are largely new to oceanography as a field. While not without risk, the potential benefits of ML methods is creating increasing interest in the field. This review has presented some of the challenges and opportunities involved when leveraging ML techniques to improve the modeling, observing, fundamental understanding, and prediction of the ocean system.

ML applications fundamentally rely on the data available for learning, and here the ocean presents a unique challenge for ML applications. The important timescales in the ocean range from seconds to millennia, with strong interactions between processes across those scales. For example, a wind gust can trigger a phytoplankton bloom. Observations are largely sparse, noisy, and unbalanced. Temporally, very few long-timescale observations exist that span more than a few decades. A general problem with models of the ocean, either ML derived or more conventional, is that the system is highly non-stationary. With climate change, the mean state and its variance are liable to change, and a model that is trained from today's data may not be general enough to accurately represent an ocean in a warmer climate. Other components of the Earth system such as land or atmospheric models, or GFD in general, also face similar challenges, but they are exsaserbated within oceanography due to the lack of spatial and temporal observational coverage.


ML offers many avenues with which the challenges listed above could be tackled. For example, with instantaneous processes (such as radiative transfer) or small spatial scale problems (for example eddy detection), a cross-validation approach with an associated independent test dataset could be fruitful. Indeed, cross-validation is widely advisable. On longer timescales, methods related to physical constraints would likely offer better results. Hybrid approaches for combining physics-driven models and ML models are becoming increasingly useful to aid the development of ocean models and to increase their computational efficiency on HPC platforms. Such `Neural Earth System Models' (NESYM\cite{irrgang2021}) can, for example, use ML for parameterization of sub-gridscale processes. Pairings of ML and conventional methods also show great promise for improving signal-to-noise ratios during training while also anchoring ML learning to a stronger physical foundation \cite{Watson2019}.

Both the field of oceanography and ML are quickly evolving, and the computational tools available to implement ML techniques are also becoming increasingly accessible. With ample enthusiasm for ML applications to address oceanographic problems, it is also important to keep in mind that ML as a field is largely not concerned with the physical sciences. Approaching ML applications with caution and care is necessary to ensure meaningful results. The importance of increasing trust in ML methods also highlights a need for collaboration between oceanographers and ML domain experts. ML as a field is developing very swiftly, and promoting collaboration can help develop methods that are tailored to also suit the needs of oceanographers.  

This review has outlined the recent advances and some remaining challenges associated with ML adoption within oceanography. As with any promising new set of methods, while there is ample opportunity, it is also worth noting that ML adoption also comes with risk. However, exploring the full potential and charting the limits of ML within oceanography is crucial and deserves considerable attention from the research community.

\section*{Acknowledgments}
MS and VB acknowledge funding from the Cooperative Institute for Modeling the Earth System, Princeton University, under Award NA18OAR4320123 from the National Oceanic and Atmospheric Administration, U.S. Department of Commerce. RL and VB acknowledge funding from the French Government's \emph{Make Our Planet Great Again} program managed by the Agence National de Recherche under the ``Investissements d'avenir'' award ANR-17-MPGA-0010.

DJ acknowledges funding from a UKRI Future Leaders Fellowship (reference MR/T020822/1). 

PD gratefully acknowledges funding from the Royal Society for his University Research Fellowship as well as the ESiWACE, MAELSTROM and AI4Copernicus under Horizon 2020 and the European High-Performance Computing Joint Undertaking (JU; grant agreement No 823988, 955513 and 101016798). The JU received funding from the European High-Performance Computing Joint Undertaking (JU) under grant agreement No 955513. The JU receives support from the European Union’s Horizon 2020 research and innovation programme and United Kingdom, Germany, Italy, Luxembourg, Switzerland, Norway.

JB acknowledges funding from the project SFE(\#2700733) of the Norwegian Research Council. Many thanks to Laurent Bertino (NERSC) for the insightful discussion about data assimilation.

The authors also wish to thank Youngrak Cho for invaluable help with Fig. 1 and 3.

\bibliographystyle{plain}
\newcommand{\newblock}{}
\bibliography{bibliography.bib}

\newpage

\appendix
\clearpage
\onecolumn
\section{List of acronyms}

\begin{longtable}{| p{.20\textwidth} | p{.80\textwidth} |} 
\toprule
\textbf{Abbreviation} & \textbf{Description} \\ \midrule
4DVar                 & 4-dimensional variational assimilation\\
AE                    & AutoEncoder \\
AI                    & Artificial Intelligence     \\
AIC                   & Akaike information criterion \\
BIC                   & Bayesian information criterion \\
ConvNet               & Convolutional Neural Network \\
DBSCAN                & Density-Based Spatial Clustering of Applications with Noise \\
DA                    & Data Assimilation\\
DL                    & Deep Learning        \\
DNN                   & Deep Neural Network \\
ECCO                  & Estimating the Circulation and Climate of the Ocean\\
EnKF                  & Ensemble Kalman filter\\
EOF                   & Empirical Orthogonal Functions \\
GAN                   & Generative Adversarial Network \\
GFD                   & Geophysical Fluid Dynamics\\
GMM                   & Gaussian Mixture Model \\
GODAE                 & Global Ocean Data Assimilation Experiment \\
GODAS                 & Global Ocean Data Assimilation System\\
GPR                   & Gaussian Process Regression \\
GPU                   & Graphical Processing Units (GPU)\\
HPC                   & High Performance Computing\\
IAI                   & Interpretable Artificial Intelligence \\
KNN                   & K Nearest Neighbors \\
LR                    & Linear Regression \\ 
MAE                   & Mean Absolute Error \\
ML                    & Machine Learning     \\
MLP                   & Multi-Layer Perceptron \\
MSE                   & Mean Square Error \\
NESYM                 & Neural Earth System Models\\
NN                    & Neural Networks \\
PCA                   & Principal Component Analysis  \\
PINN                  & Physics Informed Neural Networks \\
RF                    & Random Forest \\
RL                    & Reinforcement Learning \\
RNN                   & Recurrent Neural Network \\
RVM                   & Relevance Vector Machines  \\
SGD                   & Stochastic Gradient Descent \\
SOM                   & Self Organizing Maps \\
SVM                   & Support Vector Machines \\
SVR                   & Support Vector Regression \\
t-SNE                 & t-distributed Stochastic Neighbor Embedding\\
UMAP                  & Uniform Manifold Approximation and Projection\\
VAE                   & Variational Autoencoder \\
XAI                   & Explainable Artificial Intelligence \\
WOCE                  & World Ocean Circulation Experiment\\
WWII                  & World War two\\
\bottomrule
\label{tabl3}
\end{longtable}
\clearpage
\twocolumn
\end{document}